%% file: Article_Classification.tex
\documentclass[review]{article}
\usepackage[compact]{titlesec}

\usepackage{lineno,hyperref}
\usepackage{geometry}
\usepackage{graphicx}
\usepackage{amsthm}
\usepackage{amssymb}
\usepackage{amsmath}
\usepackage{amsfonts}
\usepackage{xcolor}
\usepackage{caption}
\usepackage{subcaption}
\usepackage{lineno}
\usepackage{array}
\usepackage{fancyhdr}
\newtheorem{proposition}{Proposition}
\newtheorem*{proposition*}{Proposition}

\newtheorem{definition}{Definition}
\newtheorem{theorem}{Theorem}
\newtheorem*{theorem*}{Theorem}
\newtheorem{corollary}{Corollary}[theorem]
\newtheorem*{corollary*}{Corollary}

\newenvironment{proofnosquare}
 {\proof}
 {\endproof}

\modulolinenumbers[5]

\usepackage{tikz}

\usetikzlibrary{angles, quotes, intersections}
\usetikzlibrary{calc}

\usepackage{physics}
\usepackage{bm}

\usepackage{float}
\usepackage{esvect}

\usepackage{tikz}

\usetikzlibrary{angles, quotes, intersections}
\usetikzlibrary{calc}

\usepackage{physics}
\usepackage{bm}
\usepackage{dsfont}

\newcommand{\algo}{\texttt{CAD }}
\newcommand{\bfa}{\textbf{a}}
\DeclareMathOperator{\Wedge}{\bigwedge}
\DeclareMathOperator{\cupop}{\cup}

\DeclareMathOperator{\dimH}{\mathrm{dim}_{\mathcal{H}}}

\newcommand{\TASLim}{
\begin{minipage}[t]{0.9\textwidth}
    \raggedright 
    \tiny
    \vspace{0pt}
    \textbf{Property.}
    This document is not to be reproduced, modified, adapted, published translated in any material form in whole    or in part nor disclosed
     to any third party without the prior written permission of Thales Alenia Space.\\
    © 2024 Thales Alenia Space all rights reserved.  \\
  \end{minipage}
}
\fancypagestyle{plain}{

  \fancyhead{}

  \fancyhead[L]{\thepage}

  \fancyhead[R]{\nouppercase{\leftmark}}

  \fancyfoot{}

   \fancyfoot[C]{\TASLim}
}

\begin{document}
\pagestyle{plain}

\title{An original classification of obscuration-free telescopes designs unfolded in two dimensions}

\author{ B. Aymard, A. Delahaye and A. Drogoul}

\maketitle
\begin{center}
 Thales Alenia Space, 5 All\'ee des Gabians, 06150 Cannes, FRANCE. 

Corresponding authors: \texttt{[benjamin.aymard,audric.drogoul]@thalesaleniaspace.com} \\

\end{center}
\begin{abstract}
In this article we propose an original classification method for unobscured imaging systems unfolded in two dimensions.
This classification is based on a study of off-axis properties, and relies on topology and algorithm of real algebraic geometry to find at least one instance by connected component of a semialgebraic set.
Our corresponding nomenclature provides intrinsic information about the system, in terms of geometry and manufacturability.
The proposed systems for each name of the nomenclature, can be used as starting points for parallel optimizations, allowing for a much more comprehensive search of an unobscured solution, given a set of specifications.
We exemplify our method on three and four mirrors imaging systems.
\end{abstract}
\raggedright

\section{Introduction}

\input{inputs/intro}

\section{Results}

\input{inputs/results}

\section{Materials and Methods}

\input{inputs/methods}

\section{Conclusion}
In this paper, we addressed the problem of off-axis unobscured telescopes classification for $N$ mirors, with $N\leq 4$, 
formulated, when the entrance pupil diameter tends to zero, as the characterization of connected components of a semialgebraic set.
We introduced a nomenclature which is consistent in the sense that two unobscured optical systems are homotopic in the set of unobscured solutions if and only if they have same name and give one representant by class which can be used as initial point of an iterative method optimizing  optical performances under the constraint that no obscuration occurs.
This mathematically proved  nomenclature can be seen as an alphabet for opticians who needs to classify intelligibly thousands of explored solutions during telescopes design.
Let us note that, as a complement of this  work, a classification method based on real algebraic geometry  has been proposed for coaxial anastigmat telescopes  \cite{onAxis:24}.
Another interesting question, with pratical application, is to estimate the volume of the connected components, following for example \cite{Lairez:19} by restricting the studied set to a bounded box.
\appendix

\section{Proofs}
\label{sec:appendix_proofs}
\begin{proof}[Proof of Proposition~\ref{propo:is_inter}]
We suppose that there exists $i,j \in [\![0,n]\!]$ non-consecutive nor equals, such that $[a_i;a_{i+1}]\cap[a_j;a_{j+1}] \neq \emptyset$ and $[b_i;b_{i+1}]\cap[b_j;b_{j+1}] = \emptyset$.\\
We note $f(t) = (\xi_k(t))_{k \in [\![0,n+1]\!]}$. Without loss of generality (via a translation, an homothety and a rotation), we assume that  $\forall t \in [0,1],\xi_i(t)=\left(\begin{matrix}0\\0\end{matrix}\right)=\xi_{i}^{(0)}$ and $\xi_{i+1}(t)=\left(\begin{matrix}0\\1\end{matrix}\right)=\xi_{i+1}^{(0)}$.\\
We note $\xi_j(t)=\left(\begin{matrix}x_1(t)\\y_1(t)\end{matrix}\right)$ and $\xi_{j+1}(t)=\left(\begin{matrix}x_2(t)\\y_2(t)\end{matrix}\right)$.

\begin{itemize}
\item 
(1) For $t\in[0,1]$ we first suppose $x_1(t)\neq x_2(t)$, in other words we suppose that the two line segments are not parallel for $t\in[0,1]$. Then the intersection point of the two lines is $I(t) = \left(\begin{matrix}x_I(t)\\y_I(t)\end{matrix}\right) = \left(\begin{matrix}0\\y_1 - \frac{y_2-y_1}{x_2-x_1}x_1\end{matrix}\right)$. \\
We introduce~: $\begin{cases}\lambda_1(t)=y_I\\
	\lambda_2(t)=\frac{x_I-x_1}{x_2-x_1}=\frac{-x_1}{x_2-x_1}\end{cases}$the coordinates the coordinates of the intersection point  on each line that is $I = (1-\lambda_1)\xi_i+\lambda_1\xi_{i+1}=(1-\lambda_2)\xi_j+\lambda_2\xi_{j+1}$, and $\phi(\lambda)=\lambda(1-~\lambda)$.\\
\phantom{5mm} Then $I(t)\in[\xi_i;\xi_{i+1}] \Leftrightarrow \lambda_1(t) \in [0,1]  \Leftrightarrow \phi(\lambda_1(t))\geqslant 0$. \\
\phantom{20mm} and $I(t)\in[\xi_j;\xi_{j+1}] \Leftrightarrow \lambda_2(t) \in [0,1]  \Leftrightarrow \phi(\lambda_2(t))\geqslant 0$.

\vspace{2mm}

Then by the \textit{Intermediate Value Theorem}: 
\vspace{-2mm}
\begin{equation*}
\quad t_0 = \min\big\{t\in[0,1],\phi(\lambda_1(t))=0 \textrm{ or } \phi(\lambda_2(t))=0\big\}<1 \quad \textrm{is well defined.}
\end{equation*}

\begin{itemize}
\item First Case: \quad $\phi\big(\lambda_1(t_0)\big)=0 \textrm{ and } \phi\big(\lambda_2(t_0)\big)\geqslant 0$, 
then $\lambda_1(t_0) \in\{0,1\}$, 
thus $I(t_0)\in\{\xi_i(t_0),\xi_{i+1}(t_0)\}$.
\item Second Case: $\phi\big(\lambda_1(t_0)\big)\geqslant0 \textrm{ and } \phi\big(\lambda_2(t_0)\big)=0$, 
then $\lambda_2(t_0) \in\{0,1\}$, 
thus $I(t_0)\in\{\xi_j(t_0),\xi_{j+1}(t_0)\}$.
\end{itemize}

\item
(2) Else $t_1=\min(\{t\in[0,1],x_1(t)=x_2(t)\})\in\;]0,1]$ is well defined:
\begin{itemize}
\item If $x_1(t_1)\!=\!x_2(t_1)>0$ then there exists $t_2\in[0,t_1[$ \; s.t. \; $x_1(t_2)>~0$ and $x_2(t_2)>\!0$ . We conclude by applying case (1) on $f_{\lvert[0,t_2]}$. \\
Same by symmetry if  $x_1(t_1)\!=\!x_2(t_1)<0$.
\item Else $x_1(t_1)=x_2(t_1)=0$, the four points are aligned.\\
\begin{itemize}
	\item[(a)] If $\left[\xi_{j}(t_1);\xi_{j+1}(t_1)\right]\cap [\xi_i^{(0)};\xi_{i+1}^{(0)}]\neq \emptyset$ then $\{\xi_j(t_1),\xi_{j+1}(t_1)\}\cap[\xi_i^{(0)};\xi_{i+1}^{(0)}]\neq\emptyset$, 
	\item[(b)] Otherwise,  $\exists t_2\in ]0,t_1[$ such that $\left[\xi_{j}(t_2);\xi_{j+1}(t_2)\right]\cap [\xi_i;\xi_{i+1}]=\emptyset$, then   (1) applies on $f_{|[0,t_2]}$.
\end{itemize}
\end{itemize}

Hence the contradiction with the fact that $\Xi_A$ and $\Xi_B$ are homotop in $\mathcal{A}_n$.
\end{itemize}

\end{proof}

\vspace{2mm}

\begin{proof}[Proof of Proposition~\ref{propo:winding_num}]
Let $J\subset[\![0,n]\!]$ and $\gamma$ as in the statement.
As $k \notin \bigcup\limits_{j\in J}\{j,j+1\}$ and $\Xi_A \in \mathcal{A}_n$, we have $a_k \notin \bigcup\limits_{j\in J}[a_j,a_{j+1}]\supset \gamma([0,1])$. 
Thus $\mathrm{Ind}_{\gamma}(a_k)$ is well defined.

For $j\in J$, let
$F_j(t)=[\xi_j(t);\xi_{j+1}(t)]$ \qquad the $j$-th flux at time $t$.\\
With $\widetilde{J^2} = \{j_1,j_2 \in J,F_{j_1}(0)\cap F_{j_2}(0)\neq\emptyset\}$, for $(j_1,j_2)\in \widetilde{J^2}$ let:
\begin{itemize}
\item$\eta_{{j_1},{j_2}}\in\mathcal{C}^0([0,1],\mathbb{R}^2)$ \quad s.t. \quad $\{\eta_{{j_1},{j_2}}(t)\}=F_{j_1}(t)\cap F_{j_2}(t)$ \;the intersections
\item$L(t)=\{\xi_j(t)\}_{j\in J} \cup \{\eta_{{j_1},{j_2}}(t)\}_{(j_1,j_2)\in \widetilde{J^2}}$ finite subset of $\mathbb{R}^2$
\end{itemize}
By Corollary~\ref{coro:near_injectivity_reduced_path} of the \textit{Unique Reduced Path Theorem}, we can without loss of generality assume that  $\gamma$ is reduced and $\gamma(0)=\gamma(1)\in L(0)$.
 Moreover we can write $\{\lambda_1<\lambda_2<...<\lambda_N\}=\gamma^{-1}(L(0)\cap \textrm{Im}(\gamma))$ as it is finite, and we have for $i\in[\![1,N-1]\!]$, $\gamma(\lambda_i)\neq\gamma(\lambda_{i+1})$.\\
Then for $t\in[0,1]$ we define :
\begin{equation*}
\Phi_t : z\in L(0)\cap\textrm{Im}(\gamma)\subset \mathbb{R}^2\mapsto \begin{cases} \xi_{j}(t) \quad &if \quad \exists j\in [\![0,n]\!],\; z = \xi_{j}(0) = a_{j} \\ 
\eta_{{j_1},{j_2}}(t) \quad &if \quad \exists {j_1},{j_2} \in J, \; z = \eta_{{j_1},{j_2}}(0) \end{cases}
\end{equation*}
For $\lambda\in [0,1]\setminus \gamma^{-1}(L(0)\cap \textrm{Im}(\gamma))$, we note $c_\lambda \in ]0,1[$ the unique real positive number such that $\gamma(\lambda)-\gamma(\lambda_i)=c_\lambda\big(\gamma(\lambda_{i+1})-\gamma(\lambda_i)\big)$ \quad with $i\in [\![1,N]\!]$ s.t. $\lambda \in \; ]\lambda_i,\lambda_{i+1}[$.\\
\noindent And we extend the definition of $\Phi_t:\bigcup\limits_{j\in J}F_j(0)\cap\textrm{Im}(\gamma)\rightarrow \bigcup\limits_{j\in J}F_j(t)$ such that
\begin{equation*}
\Phi_t : z\in\bigcup\limits_{j\in J}(F_j(0)\setminus L(0))\cap\textrm{Im}(\gamma)\mapsto \big(1-c_\lambda\big)\Phi_t\big(\gamma(\lambda_i)\big)+c_\lambda\Phi_t\big(\gamma(\lambda_{i+1})\big)
\end{equation*}
\vspace{-4mm}

\hspace{30mm} with $\lambda\in\gamma^{-1}(\{z\})$ and $i\in[\![1,N]\!]$ s.t. $\lambda\in]\lambda_i,\lambda_{i+1}[$. \\
$\Phi_t(z)$ is independant on the choice of $\lambda\in\gamma^{-1}(\{z\})$, indeed if $\exists i,i '\in[\![1,N]\!]$,\\
$z\in  ]\gamma(\lambda_i),\gamma(\lambda_{i+1})[\;\cap\;]\gamma(\lambda_{i'}),\gamma(\lambda_{{i'}+1})[$ then $\{\gamma(\lambda_{i}),\gamma(\lambda_{{i}+1})\}=\{\gamma(\lambda_{i'}),\gamma(\lambda_{{i'}+1})\}$.\\
Finally, we set $\gamma ' = \Phi_1\circ\gamma$ , we obtain $\Psi : t \in [0,1]\mapsto \Phi_t\!\circ\!\gamma(t) - \xi_k(t)$\\ as a homotopy on $\mathbb{C}\setminus\{0\}$ between $\gamma - a_k$ and $\gamma '- b_k$.\\
By \textit{Cauchy's Homotopic Theorem} (cf. Theorem~\ref{theo:cauchy_homotopy}), we obtain :
\begin{equation*}
\begin{split}
&\mathrm{Ind}_{\gamma - a_k}(0)=\mathrm{Ind}_{\gamma '- b_k}(0)\\
ie. \qquad &\mathrm{Ind}_{\gamma}(a_k)=\mathrm{Ind}_{\gamma '}(b_k)
\end{split}
\end{equation*}
The other way follows by symmetry. Hence the equivalence.
\end{proof}


\section{Used Theorems and Lemmas}
\label{sec:appendix_theorem_winding}
\subsection{Winding number}

Let $z\in\mathbb{C}$, and $\gamma :[0,1] \xrightarrow[]{\mathcal{C}^0} \mathbb{C}$ s.t. $\gamma (0) = \gamma (1)$ be a closed curve piecewise differentiable, the winding number of $z$ relative to $\gamma$ is defined as:
\begin{equation*}
\mathrm{Ind}_{\gamma}(z) = \frac{1}{2i\pi}\int_{\gamma}\frac{d\zeta}{\zeta-z}
\end{equation*}

\begin{theorem}
	\label{theo:cauchy_homotopy}
Cauchy's Homotopic Theorem \cite{Rudin:75}:

Let $z\in\mathbb{C}$ and $\gamma,\gamma'\in\mathcal{C}^0([0,1],\mathbb{C}\!\setminus\!\{z\})$, with $\begin{cases}\gamma(0)=\gamma(1)\\ \gamma'(0)=\gamma'(1)\end{cases}$\\ two homotopic continuous closed loop on $\mathbb{C}\!\setminus\!\{z\}$.
\begin{center}Then $\mathrm{Ind}_{\gamma}(z)=\mathrm{Ind}_{\gamma '}(z)$\end{center}
\end{theorem}

\subsection{Unique Reduced Path Theorem}

Let $X\subset\mathbb{R}^n$ and  $\alpha,\beta:[0,1] \xrightarrow[]{\mathcal{C}^0} X$ two paths on X, we define the terms : 

\begin{itemize}
\item $\alpha$ is said to be a \textbf{reparametrization} of $\beta$ if there exists $h:[0,1] \xrightarrow[]{\mathcal{C}^0} [0,1]$ increasing homeomorphism, \quad such that $\alpha\circ h=\beta$
\item $\beta$ is said to be \textbf{reduced} if either:\begin{itemize} \item$\beta$ is constant

\item $\forall a,b\in[0,1]$ s.t. $a<b$, \quad $\beta_{|[a,b]}$ is not a null-homotopic loop on $X$\end{itemize}
\end{itemize}

We note $\dimH(X)=1$ the Hausdorff dimension of $X$.

\vspace{2mm}

\begin{theorem*}
Let $n\in\mathbb{N}$, and $X\subset \mathbb{R}^n,\;\dimH(X)=1$. Let $\alpha:[0,1] \xrightarrow[]{\mathcal{C}^0} X$.

Then $\alpha$ is path-homotopic on $X$ to a reduced path $\beta:[0,1] \xrightarrow[]{\mathcal{C}^0} X$ that is unique up to reparametrization. \cite{Cannon:06}

Moreover there exists a homotopy between $\alpha$ and $\beta$ with image in $\textit{Im}(\alpha)$.
\end{theorem*}

\vspace{2mm}

\begin{proposition}
\label{propo:near_injectivity_reduced_path}
Let $n \in \mathbb{N}$, $I\subset[\![1,n]\!]^2$, $(a_i)_{i\in[\![1,n]\!]}\in(\mathbb{R}^2)^n$.
For $(i,j)\in I$ we note $F_{i,j}=[a_i,a_j]$ and $X = \bigcup\limits_{(i,j)\in I}\!\!F_{i,j}$.
 Let $\beta:[0,1] \xrightarrow[]{\mathcal{C}^0} X$ a non-constant reduced path.

\begin{center}
For $L\subset X$ finite, $\beta^{-1}(L)$ is finite.
\end{center}

\end{proposition}

\vspace{2mm}

\begin{proof}
Let $L\subset X$ finite. We suppose $\beta^{-1}(L)$ infinite. \\
As $\beta^{-1}(L)=\bigcup\limits_{l\in L}\beta^{-1}(\{l\})$, there exists $l \in L$ such that $\beta^{-1}(\{l\})$ is infinite. By \textit{Bolzano-Weierstrass theorem} on the compact $[0,1]$, there exits
$(t_i)_{i\in\mathbb{N}}\in~\beta^{-1}(\{l\})^\mathbb{N}$ such that $t_i\rightarrow t\in[0,1]$. By continuity $\beta(t)=l$.

Let $H = \{\eta\in\mathbb{R}^2, \exists(i,j),(i',j')\in I,\; (i,j)\neq (i',j'),\; \{\eta\}=F_{i,j}\cap F_{i',j'}\}$ be the intersection points of the line segments. 

Let $r=\min\limits_{z'\in H\setminus \{l\}}|l-z'|>0$ well defined as $H$ finite.
Each connected component of $X\setminus H$ is homeomorphic to $]0,1[$, thus simply connected. Then any reduced path on $X\setminus H$ is injective or constant.\\
As $\beta$ is reduced and non-constant, it is injective and for $i\in\mathbb{N}$ we have $\beta^{-1}(H\setminus\{l\})\cap[t_i,t_{i+1}]\neq\emptyset$. 
Thus there exists $(u_i)_{i\in\mathbb{N}}\in~\beta^{-1}(H\setminus\{l\})^\mathbb{N}$, such that $u_i\in[t_i,t_{i+1}]$.
 Finally $\forall i \in \mathbb{N}, |\beta(u_i)-\beta(t_i)|\geqslant r$, which contradicts the continuity of $\beta$.
\end{proof}

\vspace{2mm}

\begin{corollary}
	\label{coro:near_injectivity_reduced_path}
Let $n \in \mathbb{N}$, $I\subset[\![1,n]\!]^2$, $(a_i)_{i\in[\![1,n]\!]}\in(\mathbb{R}^2)^n$.
For $(i,j)\in I$ we note $F_{i,j}=[a_i,a_j]$ and $X = \bigcup\limits_{(i,j)\in I}\!\!F_{i,j}$.

Let $H = \{\eta\in\mathbb{R}^2, \exists(i,j),(i',j')\in I,\; (i,j)\neq (i',j'),\; \{\eta\}=F_{i,j}\cap F_{i',j'}\}$ be the intersection points of the line segments. Let $L=\{a_i\}_{i\in[\![1,n]\!]}\cup H$.

Let $\alpha:[0,1] \xrightarrow[]{\mathcal{C}^0} X$, with $\alpha(0)=\alpha(1)$ a non null-homotopic loop on $X$.

\noindent Then $\alpha$ is closed-loop-homotopic on $X$ to a reduced path $\beta:[0,1] \xrightarrow[]{\mathcal{C}^0} X$ such that: 

\begin{itemize}
\item $\beta(0)=\beta(1)\in H$

\item There exists $N\in\mathbb{N}$, $\lambda_1<...<\lambda_N\in[0,1]$ such that $\{\lambda_1,...,\lambda_N\}=~\beta^{-1}(L\cap\textrm{Im}(\alpha))$,
and for $i\in[\![1,N-1]\!]$, $\beta(\lambda_i)\neq\beta(\lambda_{i+1})$.
\end{itemize}

\end{corollary}

\vspace{2mm}

\begin{proof}
For $(i,j)\in I$, $\dimH([a_i,a_j])=1$. Thus by finite union $\dimH(X)=1$. 

Thus by the \textit{Unique Reduced Path Theorem}. There exists $\beta$ path-homotopic on $X$ to $\alpha$. With $f:[0,1]\rightarrow\mathcal{C}^0([0,1],X)$ the corresponding homotopy. 
The end points being fixed for $t\in[0,1]$ we have $f(t)(0)=f(0)(0)=\alpha(0)=\alpha(1)=f(0)(1)=f(t)(1)$, thus $f$ is a closed-loop homotopy.

We suppose $\beta(0)=\beta(1)\notin H$. Each connected component of $X\setminus H$ is homeomorphic to $]0,1[$, thus simply connected. As $\alpha$ and $\beta$ are not null-homotopic loop on $X$, there exists $\lambda\in[0,1]$, such that $\beta(\lambda)\in H$.\\
For $\theta\in[0,1]$, let $g_\theta:t\in[0,1]\mapsto\mathrm{mod}_1(t+\theta\lambda)$. Then $\theta\mapsto \beta\circ g_\theta$ is a closed-loop homotopy on $X$ between $g_0=\beta$ and $g_1$, and by construction $g_1(1)=g_1(0)=\lambda$.

We apply \textit{Unique Reduced Path Theorem} on $g_1$ to obtain $\beta$ reduced path closed-loop-homotopic to $\alpha$ such that $\beta(0)=\beta(1)\in H$.

Finally the existence of $\{\lambda_1,...,\lambda_N\}$ is given by Proposition~\ref{propo:near_injectivity_reduced_path}. By contradiction, if there exists $i\in[\![1,N-1]\!]$, such that $\beta(\lambda_i)=\beta(\lambda_{i+1})$ , then $\beta_{|]\lambda_i,\lambda_{i+1}[}\subset X\setminus H$. Each connected component of $X\setminus H$ is homeomorphic to $]0,1[$, thus simply connected. Then $\beta_{|[\lambda_i,\lambda_{i+1}]}$ is a null-homotopic path, which contradicts $\beta$ reduced.
\end{proof}


\section{Uniqueness of the class in 3D}
\label{sec:appendix_3D}
\vspace{2mm}

In the following section we will prove that for $n\in\mathbb{N}$ there is a unique class of unobscured systems with n mirrors if we allow the systems to unfold in three dimensions (ie. we do not impose the planar symmetry of the system).

\vspace{2mm}

Let be $\mathbf{f}:\mathbb{R}^6\to \mathbb{R}^3$ defined by $\mathbf{f}(\mathbf{x},\mathbf{y})=\mathbf{x}\wedge \mathbf{y}$ and $\mathcal{V}=\{(\mathbf{x},\mathbf{y})\in \mathbb{R}^6,\ \mathbf{f}(\mathbf{x},\mathbf{y})=0\}$.
By trivial computations we get that $$\textrm{jac}(\mathbf{f})(\mathbf{x},\mathbf{y})=\left(\begin{array}{cccccc}0&y_3&-y_2&0&-x_3&x_2\\
	-y_3&0&y_1&x_3&0&-x_1\\
	y_2&-y_1&0&-x_2&x_1&0
\end{array}\right).$$
For $(\mathbf{x},\mathbf{y})\in \mathcal{V}\backslash\{ O\}$, let us show that $\textrm{rank}(jac(\mathbf{f}))(\mathbf{x},\mathbf{y})\leq 2$.
Indeed, 
\[x_2\left(\begin{array}{ccc}0&y_3&-y_2\\
	-y_3&0&y_1\\
	y_2&-y_1&0
\end{array}\right)=y_2\left(\begin{array}{ccc}0&-x_3&x_2\\
	x_3&0&-x_1\\
-x_2&x_1&0
\end{array}\right)
	\]
	and 
	\[x_2\left(\begin{array}{ccc}0&-x_3&x_2
	\end{array}\right)
	+x_2\left(\begin{array}{ccc}
		x_3&0&-x_1
	\end{array}\right)
	=x_3\left(\begin{array}{ccc}
	-x_2&x_1&0
	\end{array}\right).
		\]
Besides, if $(\mathbf{x},\mathbf{y})\neq O$, it is easily checked that $\textrm{rank}(jac(\mathbf{f}))(\mathbf{x},\mathbf{y})\geq 2$.
We deduce that the set of sigular points of  $\mathcal{V}$ are given by 
\begin{align*}
	\textrm{sing}(\mathcal{V})&=\{O\}
\end{align*}
	Hence $\textrm{dim}(\textrm{sing}(\mathcal{V}))=0$ and $\textrm{dim}(\mathcal{V}\backslash \{O\})=4$.
	We deduce $\mathcal{V}$ is of codimension 2 and  that $\mathcal{E}=\mathbb{R}^6\backslash\mathcal{V}$ is connected thus path connected.\\
	Similarly, 
	\begin{align*}
		\mathrm{Let}\quad\mathcal{E}_n&=\mathbb{R}^{3(n+1)}\backslash \mathcal{V}_n\\
		\mathrm{with }\quad \mathcal{V}_{n}&=\cupop\limits_{\substack{i,j\in[\![0,n]\!]\\i<n\\j\notin \{i,i+1\}}}\mathcal{V}_{i,j,n}\\
		\mathrm{with }\quad \mathcal{V}_{i,j,n}&=\left\{(x_0,...,x_n)\in(\mathbb{R}^3)^{n+1},\mathbf{f}(x_j-x_i,x_j-x_{i+1}) = 0\right\}\\
	\end{align*}
	From the previous proof, we deduce that $\textrm{dim}(\mathcal{V}_{i,j,n}\backslash\{O\})  \leq 3n+1$ for all $ i,j\in[\![0,n]\!]$, hence by finite union $\textrm{dim}(\mathcal{V}_{n}\backslash\{O\})\leq 3n+1$ and $\mathcal{E}_n$ is connected thus path connected.
		

\end{document}

%% file: inputs/intro.tex
In the framework of optical design, the recent field of freeform optics, 
based on the relaxation of the rotational symmetry constraint,
has given access to a broader space of solutions.
Freeform designs are often more compact for the same specifications in terms of optical performances.
However, the number of degrees of freedom has soared, and so has the number of possible configurations.
This is a key difference compared to historical designs, for which only few configurations were possible, solutions of a set of low order polynomial equations.

A widely-used method when designing freefrom imaging systems consists of three steps \cite{Rolland:18}.
First, provide an on-axis initialization, using for instance the Korsch equations \cite{Korsch:91}, 
defining the curvatures and conic constants of every surface.
Solutions of Korsch equations ensure that the initial system respects several constraints:
focal length, back focal length, Petzval curvature condition, and suppression of third order Seidel aberrations. 
Second step, the use of surface tilts, together with pupil off-axis and field off-axis, opens the system in order to suppress all obscurations 
(central obstruction, vignetting) \cite{Gross:19}. 
This process introduces optical aberrations, that have to be corrected.
Finally, the third step consists of using orthogonal polynomials corrections (such as Zernike basis), or local polynomial corrections (for instance using NURBS surfaces \cite{Chrisp:16}),
to suppress optical aberrations.

When looking for a design which meets specifications, the optical designer can follow the latter procedure, 
which produces one design, within a brute force search exploration of the space of parameters.
A design can then be selected, as the best configuration among all the computed solutions. 
However, there is no certainty that the found solution is the best possible. 
A different discretization in the brute force search could lead to another solution, possibly dramatically different.
Not only several solutions may co-exist: they could qualitatively behave very differently, 
either geometrically (volume, obscuration), optically (WFE, distorsion), or from the manufacturability point of view (sag, local slope).

This type of approach has recently been sucessfully applied to study the case of three mirrors \cite{Zhu:21} imaging systems.
The volume of results being large, the authors had to sort the solutions, and, in doing so, proposed a classification.
We noticed several couples of design, that, despite being very similar, both in geometry and performance, were far apart in the given classification
(for instance the couple L14-1 and L35-1, or the couple L73-2 and L18-1).
This classification attributes a unique name to each system, but it does not carry intrinsic properties, 
describing instead the behavior of the given system. 
A designer cannot retrieve optical information about the system from its code.

A similar study, also based on a brute force search, was carried out in the case of four mirrors imaging systems \cite{Papa:21}.
Another way of sorting the results was proposed: the designs were sorted on a two dimensional array, 
with the F-number in abscissa, and the Field Of View (FOV) in ordinates. 
With this approach one can see, depending on the specifications, which configuration is the best, among the result of the brute force search
(and therefore, the result is directly linked with the resolution of the explored space).
But the same critique applies: in this case there is no exhaustive unambiguous classification,
and therefore there is no way to guess the behavior of one of them before actually trying it.

Interestingly, both of these references were confronted to the classification problem, that naturally arises during design exploration.
Given the large number of solutions (in practice of the order of hundred of thousands), the optical designer can not reasonably review them one by one.
There is a need for a classification, in order to understand and identify classes of solutions sharing qualitative properties.
Several forms of classification have been proposed in the litterature.
In \cite{Howard:00}, the authors propose to class systems according to the sign of the tilt of the mirrors.
This method was later used in \cite{Papa:18}, to propose a list of 15 three miurrors configurations unfolded in two dimensions.
A different approach was followed in the thesis \cite{Houllier:21}, where the author presents a classification of folding geometry, 
by giving a name describing intuitively the general shape of the optical flux:  Z, 4, W and $\gamma$.
However, none of the proposed method has been proven to be exhaustive, and, for instance, 
the exhaustive list of three mirrors telescope un folded in two dimensions has not been unveiled, to this day, to the best of the authors knowledge.
The aim of this article is to find a way to sort the results, group them by classes sharing similar intrinsic propeties (geometry, optics, manufacturability),
find a way to name those classes, and prove that the corresponing classification is exhaustive.

Historically, the introduction of classification has often led to major breakthroughs.
The most prominent example is the classification of living organisms, that led to the Darwinian theory of evolution.
But another example, close to the methodology developed in this article, also came during the 19th century.
August Ferdinand Moebius, at the time working on the problem of classification of surfaces, 
introduced a nomenclature, allowing a complete topological description of surfaces in alphabetical terms \cite{Moebius:1863}.
This result led to the complete clasification of compact surfaces in three dimensions.
Around the same period of time, the scientific community tried to classify all the knots, and a first table was proposed by Tait.
However, in this case, the right theoretical framework (knot polynomials) came later, and so did the nomenclatures (for instance, the famous Dowker notation \cite{Dowker:83}).
Our classification and its corresponding nomenclature aims to emulate the work of Moebius to describe an imaging system in terms of an alphabet.
With it, like a musician who can read or writes music, and understand without actually playing the score, 
an optical designer can write and read optical designs unfolded in two dimensions, without tracing them.

Our classification strategy relies on expressing the problem as the study of the connected components of a semi algebraic set.
Within this framework, computer algebra algorithms, such as Cylindrical Algebraic Decomposition (\algo, see \cite{Collins:75}) are used to complete mathematical proofs.
Let us note that similar approaches have already been used in other domains,
such as robotics \cite{CapcoEldin:20} or Biology \cite{chauvin:00}. 
For example, in \cite{CapcoEldin:20}, authors address the problem of the number of connected components of 
the complementary of the kinematic singularity locus of a robot, written as a Zariski open set.

This article is organized as follows.
First, we introduce the notion of homotopically equivalent unobscured optical systems,  
and we identify topological invariants.
Then we provide a nomenclature, allowing a univoque description of optical system, for two, three, and four mirors,
we prove rigorously that the latter is exhaustive.
Eventually, we propose a method to identify at least one element by class.
Combining the two approaches, one leading to a upper bound in the number of classes, and one giving a lower bound, 
we identify all the classes of telescopes, for two to four mirrors configurations.

%% file: inputs/results.tex
\subsection{Topological invariants}

We model the optical system by a succession of line segments (representing the optic axis).
 With $a_0$ the position of the object, $(a_1,...,a_n)$ the positions of the mirrors and $a_{n+1}$ the position of the image, 
 an optical system with planar symmetry can be represented as : $(a_0,...,a_{n+1})\in(\mathbb{R}^2)^{n+2}$.

\vspace{2mm}

For $a,b\in\mathbb{R}^2$ we note $[a;b]=\bigl\{(1-\lambda)a+\lambda b \; , \lambda \in [0,1]\bigr\}$ \\
\phantom{\qquad\qquad\qquad\qquad\quad}and $(a;b)=\bigl\{(1-\lambda)a+\lambda b \; , \lambda \in \mathbb{R}\bigr\}$

\vspace{2mm}

In this representation the system is \textit{obscured} when the object, the image or a mirror $a_j$ passes through a flux $[a_i;a_{i+1}]$. The system is \textit{grazing} when a mirror is parallel with the flux $\,\,\,  i.e. \quad a_{k+1}\in [a_k;a_{k+2}]$ .

As two optical systems are considered of the same family when we can continuously go from one to the other without obscuration, the families correspond exactly to the path connected components of the following set. 

For $x,y,z,t\in\mathbb{R}$ we note $\left(\begin{matrix}x\\y\end{matrix}\right)\wedge\left(\begin{matrix}z\\t\end{matrix}\right) = xz-yt$ \quad .

\begin{definition}

For $n\in\mathbb{N}$ and $\bfa=(a_0,...,a_{n+1})\in(\mathbb{R}^2)^{n+2}$ let us define the  cost function

\begin{equation}
	\label{eq:obscRazCost}
\mathcal{J}(\bfa)=\sum\limits_{\substack{0\leqslant i\leqslant n \\ 0\leqslant j\leqslant n+1\\j \notin \{i-1,i,i+1,i+2\}}}\mathds{1}_{[a_i;a_{i+1}]}(a_j) \;\; + \sum\limits_{0\leqslant k\leqslant n-1 }\mathds{1}_{(a_k;a_{k+1})}(a_{k+2}) \quad ,
\end{equation}

and

\begin{equation*}
\hspace{-8mm}
\mathcal{A}_n = \left\{ \bfa\in(\mathbb{R}^2)^{n+2}, \; \mathcal{J}(\bfa)=0
\right\} \quad .
\end{equation*} 
\begin{figure}[H]
    \centering
	\begin{subfigure}[c]{0.3\textwidth}
		\centering
		\includegraphics[width=\textwidth]{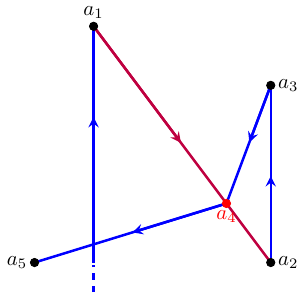}
        \caption{\label{fig:obscuration_example_fig}}
        
	\end{subfigure}
    \hspace{5mm}
	\begin{subfigure}[c]{0.3\textwidth}
		\centering
		\includegraphics[width=\textwidth]{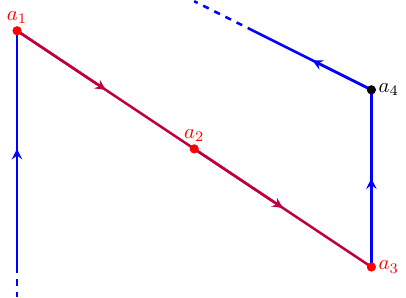}
        \caption{\label{fig:grazing_example_fig}}
        
	\end{subfigure}
	\caption{\label{fig:cas_obscure_rasant}Non admissible cases according to $\mathcal{J}$ \eqref{eq:obscRazCost}. With (a) for obscuration and (b) for grazing.}
    \label{fig:non_admissible_cases}
\end{figure}
\vspace{1.2mm}

Let us denote $C\!\!\!C(\mathcal{A}_n)$ these path connected componants, thus $\mathcal{A}_n = \bigsqcup\limits_{C\in C\!\!\!C(\mathcal{A}_n)}\!\!\!\!C$.

\end{definition}

\vspace{5mm}
Let us comment the expression of the obscuration cost $\mathcal{J}$ \eqref{eq:obscRazCost}. Its first term is not null if there exists a mirror obscuring the $i$-th flux (see Figure~\ref{fig:cas_obscure_rasant}-(a)) and non adjacent to it;  the second term
penalizes aligned consecutive triplets that is, it is not null if either the $k+2$-th mirror  obscurs the $k$-th flux or the $k+1$-th mirror is grazing (see Figure~\ref{fig:cas_obscure_rasant}-(b)).

It is important to note that the connected componants of $\mathcal{A}_n$ are not trivial. For example they are not convex (even with only one mirror):
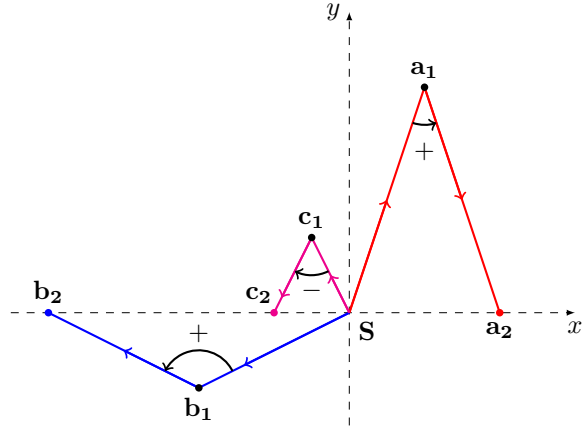
\begin{figure}[H]
\centering
\begin{tikzpicture}
		
		\coordinate (O) at (0,0);
		\coordinate (A) at (1,3);
		\coordinate (B) at (2,0);
		\coordinate (A') at (-2,-1);
		\coordinate (B') at (-4,0);
		\coordinate (A'') at (-.5,1);
		\coordinate (B'') at (-1,0);

		\draw (0,0) node [below right] {$\vb{S}$};
		\draw (1,3) node [above] {$\vb{a_1}$};
		\draw (2,0) node [below] {$\vb{a_2}$};
		\draw (-2,-1) node [below] {$\vb{b_1}$};
		\draw (-4,0) node [above] {$\vb{b_2}$};
		\draw (-.5,1) node [above] {$\vb{c_1}$};
		\draw (-.9,0) node [above left] {$\vb{c_2}$};
		
		\draw[ultra thin, dashed,  -latex] (-4.5,0) -- (3,0) node [below] {$x$};
		\draw[ultra thin, dashed,  -latex] (0,-1.5) -- (0,4) node [left] {$y$};
		
		\draw[thick, red]      (O)  -- (A);
		\draw[thick, red, ->]  (O)  -- ($.5*(A)$);
		\draw[thick, red]      (A)  -- (B);
		\draw[thick, red, ->]  (A)  -- ($.5*(A)+.5*(B)$);

		\draw[thick, blue]     (O)  -- (A');
		\draw[thick, blue, ->] (O)  -- ($.7*(A')$);
		\draw[thick, blue]     (A') -- (B');
		\draw[thick, blue, ->] (A') -- ($.5*(A')+.5*(B')$);

		\draw[thick, magenta]      (O)  -- (A'');
		\draw[thick, magenta, ->]  (O)  -- ($.5*(A'')$);
		\draw[thick, magenta]      (A'')-- (B'');
		\draw[thick, magenta, ->]  (A'')-- ($.2*(A'')+.8*(B'')$);
		
		\foreach \n in {A,A',A''}
			\node at (\n)[circle,fill, inner sep=1pt]{};

		\node at (B)[circle, red, fill, inner sep=1pt]{};
		\node at (B')[circle, blue, fill, inner sep=1pt]{};
		\node at (B'')[circle, magenta, fill, inner sep=1pt]{};
					
		\pic[draw, ->, thick, "$+$", angle eccentricity=1.7] {angle = O--A--B};
		\pic[draw, ->, thick, "$+$", angle eccentricity=1.45] {angle = O--A'--B'};
		\pic[draw, <-, thick, "$-$", angle eccentricity=1.4] {angle = B''--A''--O};
		
\end{tikzpicture}
\caption{\centering Three systems composed of the source $S=a_0=b_0=c_0=(0,0)$ , a mirror at $a_1$ (resp. $b_1$, $c_1$) and the image at $a_2$ (resp. $b_2$, $c_2$). The third system is the mean of the first and second system  ie. $(c_0,c_1,c_2)=\frac{1}{2}(a_0,a_1,a_2)+\frac{1}{2}(b_0,b_1,b_2)$  Yet it does not share the same connected component (as shown by the sense of rotation, cf. Proposition \ref{propo:sense_rota}.)}
\end{figure}

\vspace{5mm}
To investigate the connected components of $\mathcal{A}_n$, we first study those of the following set $\mathcal{B}_n$:

\begin{definition}
For  $n\in\mathbb{N}$ and $\bfa=(a_0,...,a_{n+1})\in(\mathbb{R}^2)^{n+2}$ let us define

\begin{equation*}
\mathcal{K}(\bfa)=\sum\limits_{\substack{0\leqslant i\leqslant n \\ 0\leqslant j\leqslant n+1\\j \notin \{i,i+1\}}}\mathds{1}_{(a_i;a_{i+1})}(a_j) \quad .
\end{equation*}
and
\begin{equation*}
\mathcal{B}_n = \left\{ \bfa\in(\mathbb{R}^2)^{n+2}, \mathcal{K}(\bfa)=0\right\} \quad .
\end{equation*}
\end{definition}

\vspace{5mm}

There is an inequality between the number of connected components of $\mathcal{A}_n$ and $\mathcal{B}_n$ which is easily demonstrated when we consider the topological properties of the latters.

\begin{proposition}
\label{propo:AB_open_alge}
For $n\in\mathbb{N}$, 
$\mathcal{A}_n,\mathcal{B}_n$ are dense open semialgebraic subsets of $(\mathbb{R}^2)^{n+2}$, with $\mathcal{B}_n \subset \mathcal{A}_n$ and $\mathcal{B}_n$ \textit{Zariski open} \cite{Cox:15}.\\
More precisely, $\mathcal{B}_n$ and $\mathcal{A}_n$ write as :
\begin{align*}
\mathcal{B}_n &= \biggl\{\bfa \in (\mathbb{R}^2)^{n+2},\ \prod\limits_{\substack{0\leqslant i\leqslant n \\ 0\leqslant j\leqslant n+1\\j \notin \{i,i+1\}}} \!\hat{P}(a_i,a_{i+1},a_j) \neq 0\biggr\},\\
\mathcal{A}_n &= \biggl\{\bfa \in (\mathbb{R}^2)^{n+2},\ \Wedge\limits_{\substack{0\leqslant i\leqslant n \\ 0\leqslant j\leqslant n+1\\j \notin \{i-1,i,i+1,i+2\}}} \!\left(\hat{P}(a_i,a_{i+1},a_j) \neq 0\right)\vee \left(\hat{R}(a_i,a_{i+1},a_j) < 0 \right)\vee \left(\hat{R}(a_{i+1},a_{i},a_j) < 0 \right)\\
&\quad \quad \text{ and } \Wedge\limits_{0\leqslant k\leqslant n-1 } \left(\hat{P}(a_k,a_{k+1},a_{k+2})\neq 0\right)\biggr\}
\end{align*}
with, by denoting $a_i=\left( x_i, y_i\right)$:
\begin{equation*}
	\begin{aligned}
	\hat{P}(a_1,a_2,a_3) &= x_1y_2-x_1y_3+x_2y_3-x_2y_1+x_3y_1-x_3y_2  \in\mathbb{Q}[x_1,x_2,x_3,y_1,y_2,y_3],\\
	\hat{R}(a_1,a_2,a_3) &= (x_2-x_1)(x_3-x_1)+(y_2-y_1)(y_3-y_1)\in\mathbb{Q}[x_1,x_2,x_3,y_1,y_2,y_3].
	\end{aligned}
\end{equation*}

\end{proposition}

\begin{proof}[Proof of Proposition~\ref{propo:AB_open_alge}]
For $i,j,k\in [\![0,n+1]\!]$ we have $a_j \in [a_i;a_{i+1}] \Rightarrow a_j \in (a_i;a_{i+1})$, and $a_{k+1} \in [a_k;a_{k+2}] \Rightarrow a_j \in (a_i;a_{i+1})$ with $\begin{cases}i=k\\j=k+1\end{cases}$\!, \\hence the inclusion.

Then with $a_k = \left(\begin{matrix} x_k \\ y_k \end{matrix}\right) \in \mathbb{R}^2 $, the condition $a_j\notin (a_i;a_{i+1})$ can be written as:

\begin{equation*}
\left(\begin{matrix} x_j-x_i \\ y_j-y_i \end{matrix}\right) \wedge \left(\begin{matrix} x_j-x_{i+1} \\ y_j-y_{i+1} \end{matrix}\right) \neq 0 
\end{equation*}
\begin{equation*}
ie. \quad  (x_j-x_i)(y_j-y_{i+1})-(x_j-x_{i+1})(y_j-y_i) \neq 0 
\end{equation*}
\begin{equation*}
ie. \quad  \hat{P}(x_i,x_{i+1},x_j,y_i,y_{i+1},y_j) \neq 0 
\end{equation*}

We obtain $\mathcal{B}_n $ as Zariski open (thus semialgebraic), and a dense open subset of $\mathbb{R}^{2(n+2)}$.

\vspace{-2mm}

\begin{equation*}\hspace{5mm}\textrm{Finally :}
\quad  a_k\in [a_i;a_j]  \; \Leftrightarrow \;\begin{cases}a_k\in (a_i;a_j) \\ \vv{a_ia_j}\cdot \vv{a_ia_k} \geqslant 0 \\ \vv{a_ja_i}\cdot \vv{a_ja_k} \geqslant 0 \end{cases} \Leftrightarrow \; \begin{cases}\hat{P}(x_i,x_j,x_k,y_i,y_j,y_k)=0  \\ \hat{R}(a_i,a_j,a_k) \geqslant 0 \\ \hat{R}(a_j,a_i,a_k)  \geqslant 0\end{cases} 
\end{equation*}



%
Hence writing 
\begin{align*}
	\mathcal{A}_n &= \biggl\{\bfa \in (\mathbb{R}^2)^{n+2},\ \Wedge\limits_{\substack{0\leqslant i\leqslant n \\ 0\leqslant j\leqslant n+1\\j \notin \{i-1,i,i+1,i+2\}}} \!\left(a_j\not\in [a_i;a_{i+1}]\right)\text{ and }  \Wedge\limits_{0\leqslant k\leqslant n-1 } \left(a_{k+2}\not\in (a_k;a_{k+1})\right)\ \biggr\}\\	
&= \biggl\{\bfa \in (\mathbb{R}^2)^{n+2},\ \Wedge\limits_{\substack{0\leqslant i\leqslant n \\ 0\leqslant j\leqslant n+1\\j \notin \{i-1,i,i+1,i+2\}}} \!\left(a_j\not\in (a_i;a_{i+1})\right)\vee \left( \vv{a_ia_{i+1}}\cdot \vv{a_ia_j} < 0\right)\vee \left( \vv{a_{i+1}a_i}\cdot \vv{a_{i+1}a_j} < 0\right)\\
&\quad \quad \text{ and }  \Wedge\limits_{0\leqslant k\leqslant n-1 } a_{k+2}\not \in (a_k;a_{k+1})\biggr\}
\end{align*}
we obtain  $\mathcal{A}_n$ as a semialgebraic supset of $\mathcal{B}_n$,  dense in $\mathbb{R}^{2(n+2)}$.

\end{proof}

\vspace{2mm}

\begin{proposition}
\label{propo:AB_ineq}

$\quad \forall C \in C\!\!\!C(\mathcal{A}_n) , \exists C' \in C\!\!\!C(\mathcal{B}_n) , C' \subset C $

\vspace{-4.5mm}

\begin{gather*}
Thus  \quad \lvert C\!\!\!C(\mathcal{A}_n) \rvert \leqslant \lvert C\!\!\!C(\mathcal{B}_n) \rvert \quad .
\end{gather*}

\end{proposition}

\vspace{3mm}

\begin{proof}[Proof of Proposition~\ref{propo:AB_ineq}]
For $C \in C\!\!\!C(\mathcal{A}_n)$ , C is a non empty open set of $(\mathbb{R}^2)^{n+2}$ (as an open set of the open set $\mathcal{A}_n$), and $\mathcal{B}$ is dense in $(\mathbb{R}^2)^{n+2}$. \\
Thus $C \cap \mathcal{B}_n \neq \emptyset$ , and we obtain $C'\in C\!\!\!C(\mathcal{B}_n)$ such that $C \cap C' \neq \emptyset$. \\
Moreover $C \subset \mathcal{A}_n \subset \mathcal{B}_n$ and $C$ path-connected, thus $C \subset C'$ by definition of a connected component of $\mathcal{B}_n$ .
\end{proof}

\vspace{5mm}

In the rest of this section, let $\Xi_A=(a_0,...,a_{n+1}) , \Xi_B=(b_0,...,b_{n+1})\in \mathcal{A}_n$  be two systems with $n$ mirrors of the same path connected components.
 Hence there exists an homotopy $f \in \mathcal{C}([0,1],\mathcal{A}_n) $ such that $f(0)=\Xi_A$ and $f(1)=\Xi_B$.\\
  We note for $t\in[0,1]$, $f(t) = (\xi_i(t))_{i\in[\![0,n+1]\!]}\in (\mathbb{R}^2)^{n+2}$.\\
  We note for $\bfa=(a_0,...,a_{n+1})\in\mathbb{R}^{2(n+2)}$ and $(i,j,k)\in[\![0,n+1]\!]^3$, $P_{(i,j,k)}(\bfa)=\hat{P}(a_i,a_j,a_k)$.

\vspace{5mm}

In the next four propositions, we exhibit topological invariants that will enable us to determine if two systems are not in the same path connected components. 
Topological invariants are 
\begin{definition}
	\label{de:topologicalInvariant}
	Let $X$ be a topological space and $N$ be a countable set, we say that $\mathcal{S}:X\to N$  is a topological invariant over $X$ if 
	\[ \forall (x,x')\in X^2\ [\exists \gamma \in C^0([0,1],X)\ : \  \gamma(0)=x\land \gamma(1)=x']\Longrightarrow \mathcal{S}(x)=\mathcal{S}(x')
		\]
	We say that $\mathcal{S}$ is exact if we replace the implication by an equivalence.
 \end{definition}
 In the sequel, for simplicity, we will define topological invariants as properties rather than through applications. 
 We only give the final topological invariant (see subsection~\ref{subsec:goffa}), gathering the next four propositions.
\begin{proposition}
\label{propo:sense_rota}
For $i\in[\![0,n-1]\!]$, we have : 
\begin{equation*}
(a_{i+1}-a_i)\wedge(a_{i+2}-a_{i+1}) >0\quad\Leftrightarrow\quad(b_{i+1}-b_i)\wedge(b_{i+2}-b_{i+1})>0 \quad .
\end{equation*}

In other words the sense of rotation of consecutive triplets is preserved.
\end{proposition}

\vspace{2mm}
    
\begin{proof}[Proof of Proposition~\ref{propo:sense_rota}]
We suppose that $\exists i \in [\![0,n-1]\!] , \begin{cases}P_{(i,i+1,i+2)}(\Xi_A) < 0 \\ P_{(i,i+1,i+2)}(\Xi_B) > 0 \end{cases}$. 
By the \textit{Intermediate Value Theorem}, there exists $t_0 \in [0,1]$ such that \\ $P_{(i,i+1,i+2)}(f(t_0)) = 0$, thus by definition $f(t_0) \notin \mathcal{A}_n$, \\ hence the contradiction.
\end{proof}

\vspace{5mm}

\begin{proposition}
\label{propo:is_inter}
For $i,j\in[\![0,n]\!]$ neither equal nor consecutive, we have : 
\begin{equation*}
[a_i;a_{i+1}]\cap[a_j;a_{j+1}] = \emptyset\quad\Leftrightarrow\quad[b_i;b_{i+1}]\cap[b_j;b_{j+1}]=\emptyset \quad .
\end{equation*} 
\end{proposition}

\begin{proofnosquare}[Proof of Proposition~\ref{propo:is_inter}]
See Appendix~\ref{sec:appendix_3D}.
\end{proofnosquare}

\vspace{5mm}

\begin{proposition}
\label{propo:sign_inter}
For $i,j\in[\![0,n]\!]$ neither equal nor consecutive, \\
s.t. $[a_i;a_{i+1}]\cap[a_j;a_{j+1}] \neq \emptyset$ (thus $[b_i;b_{i+1}]\cap[b_j;b_{j+1}] \neq \emptyset$ by Proposition~\ref{propo:is_inter}.).

\vspace{2mm}

We have : 
\begin{equation*}
(a_{i+1}-a_i)\wedge(a_{j+1}-a_j) >0\quad\Leftrightarrow\quad(b_{i+1}-b_i)\wedge(b_{j+1}-b_j)>0 \quad .
\end{equation*} 
\end{proposition}
\begin{proof}
	We have by Proposition~\ref{propo:is_inter}:  $ \forall t \in [0,1], [\xi_i(t),\xi_{i+1}(t)]\cap[\xi_j(t),\xi_{j+1}(t)]\neq\emptyset$, thus $\Bigl((\xi_{i+1}(t)-\xi_i(t))\wedge(\xi_{j+1}(t)-\xi_j(t))=0 \Bigr)\Rightarrow \Bigl(f(t)\notin {\mathcal{A}_n}\Bigr)$, hence the result by the \textit{Intermediate Value Theorem} and the continuity of $\wedge$.

\end{proof}

\vspace{2mm}

\begin{proofnosquare}[Proof of Proposition~\ref{propo:sign_inter}]
See Appendix~\ref{sec:appendix_proofs}.
\end{proofnosquare}

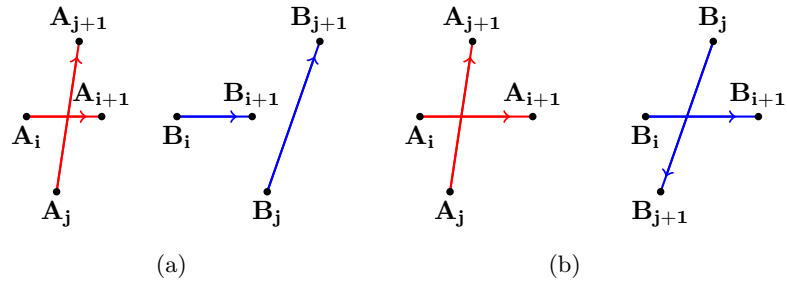
\begin{figure}[H]
     \centering
     \begin{subfigure}[b]{0.3\textwidth}
         \centering

		 \begin{tikzpicture}
		
			\coordinate (A) at (-1,0);
			\coordinate (B) at (0,0);
			\coordinate (A') at (-2,0);
			\coordinate (B') at (1,0);
			\coordinate (A'') at (-1.3,1);
			\coordinate (B'') at (1.9,1);
			\coordinate (A''') at (-1.6,-1);
			\coordinate (B''') at (1.2,-1);
	
			\draw (-1,0)    node [above] {$\vb{A_{i+1}}$};
			\draw (0,0)     node [below] {$\vb{B_i}$}    ;
			\draw (-2,0)    node [below] {$\vb{A_i}$}    ;
			\draw (1,0)     node [above] {$\vb{B_{i+1}}$};
			\draw (-1.3,1)  node [above] {$\vb{A_{j+1}}$};
			\draw (1.9,1)   node [above] {$\vb{B_{j+1}}$};
			\draw (-1.6,-1) node [below] {$\vb{A_j}$}    ;
			\draw (1.2,-1)  node [below] {$\vb{B_j}$}    ;
	
			\draw[thick, red, -]    (A)  -- (A')  ;
			\draw[thick, red, -]    (A'')-- (A''');
			\draw[thick, blue,-]    (B)  -- (B')  ;
			\draw[thick, blue,-]    (B'')-- (B''');
			
			\draw[thick, red, ->]    (A')   -- ($.8*(A)+.2*(A')$)    ;
			\draw[thick, red, ->]    (A''') -- ($.9*(A'')+.1*(A''')$);
			\draw[thick, blue,->]    (B)   -- ($.2*(B)+.8*(B')$)     ;
			\draw[thick, blue,->]    (B''') -- ($.9*(B'')+.1*(B''')$);
			
			\foreach \n in {A,A',A'',A''',B,B',B'',B'''}
				\node at (\n)[circle,fill, inner sep=1pt]{};
		
	\end{tikzpicture}
         \caption{}
         \label{fig:is_inter_fig}
	 \end{subfigure}
	 \hspace{5mm}
     \begin{subfigure}[b]{0.3\textwidth}
         \centering

		 \begin{tikzpicture}
		
			\coordinate (A) at (-.5,0);
			\coordinate (B) at (1,0);
			\coordinate (A') at (-2,0);
			\coordinate (B') at (2.5,0);
			\coordinate (A'') at (-1.3,1);
			\coordinate (B'') at (1.9,1);
			\coordinate (A''') at (-1.6,-1);
			\coordinate (B''') at (1.2,-1);
	
			\draw (-.5,0)    node [above] {$\vb{A_{i+1}}$};
			\draw (1,0)     node [below] {$\vb{B_i}$}    ;
			\draw (-2,0)    node [below] {$\vb{A_i}$}    ;
			\draw (2.5,0)     node [above] {$\vb{B_{i+1}}$};
			\draw (-1.3,1)  node [above] {$\vb{A_{j+1}}$};
			\draw (1.9,1)   node [above] {$\vb{B_j}$};
			\draw (-1.6,-1) node [below] {$\vb{A_j}$}    ;
			\draw (1.2,-1)  node [below] {$\vb{B_{j+1}}$}    ;
	
			\draw[thick, red, -]    (A)  -- (A')  ;
			\draw[thick, red, -]    (A'')-- (A''');
			\draw[thick, blue,-]    (B)  -- (B')  ;
			\draw[thick, blue,-]    (B'')-- (B''');
			
			\draw[thick, red, ->]    (A')   -- ($.8*(A)+.2*(A')$)    ;
			\draw[thick, red, ->]    (A''') -- ($.9*(A'')+.1*(A''')$);
			\draw[thick, blue,->]    (B)   -- ($.2*(B)+.8*(B')$)     ;
			\draw[thick, blue,->]    (B'') -- ($.9*(B''')+.1*(B'')$);
			
			\foreach \n in {A,A',A'',A''',B,B',B'',B'''}
				\node at (\n)[circle,fill, inner sep=1pt]{};	
			
	\end{tikzpicture}
	
         \caption{}
         \label{fig:sign_inter_fig}
     \end{subfigure}
        \caption{\centering figure~\ref{fig:is_inter_fig} (resp. figure~\ref{fig:sign_inter_fig}) represents parts of two systems that are, by Proposition~\ref{propo:is_inter} (resp. Propostion~\ref{propo:sign_inter}). not in the same path connected component of $\mathcal{A}_n$.}
        \label{fig:first_two_invariants}
\end{figure}

\vspace{4mm}

\begin{proposition}
\label{propo:winding_num}

With $x,y \in \mathbb{R}$ and $\gamma:[0,1] \xrightarrow[]{\mathcal{C}^0}\mathbb{R}^2 = \left(\begin{matrix}\gamma_x\\\gamma_y\end{matrix}\right)$ s.t. $\gamma (0) = \gamma (1)$, \\
we extend the winding number notation  from complex analysis (cf. appendix): \\

\vspace{-7mm}

\begin{equation*}
\mathrm{Ind}_{\gamma}\left(\begin{matrix}x\\y\end{matrix}\right) = \mathrm{Ind}_{\gamma_x+i\gamma_y}(x+iy), \qquad\qquad  \textrm{with} \qquad i^2=-1.
\end{equation*} 

\vspace{2mm}

For $J \subset [\![0,n]\!]$ and $k \in [\![0,n+1]\!]\setminus \bigcup\limits_{j\in J}\{j,j+1\}$. We have : 

\begin{equation*}
\hspace{-8mm}
\left( \begin{aligned}\exists \gamma :[0,1] &\xrightarrow[]{\mathcal{C}^0} \bigcup\limits_{j\in J} [a_j,a_{j+1}] \\ \gamma (0) &= \gamma (1) \end{aligned}, \mathrm{Ind}_{\gamma}(a_k) = 1 \right) 
\Leftrightarrow
\left( \begin{aligned}\exists \gamma ' :[0,1] &\xrightarrow[]{\mathcal{C}^0} \bigcup\limits_{j\in J} [b_j,b_{j+1}] \\ \gamma ' (0) &= \gamma ' (1) \end{aligned}, \mathrm{Ind}_{\gamma '}(b_k) = 1 \right) .
\end{equation*} 
\end{proposition}

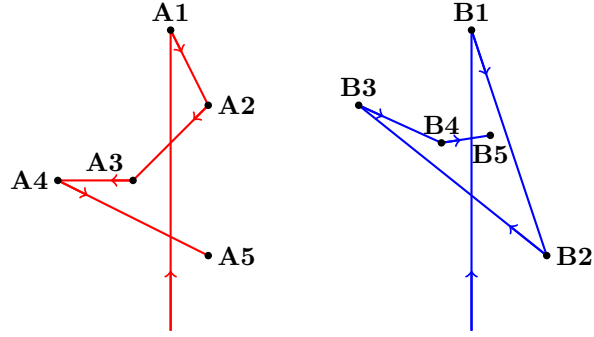
\begin{figure}[H]
\centering
{
\begin{tikzpicture}
		
		\coordinate (A) at (0,-3);
		\coordinate (B) at (0,1);
		\coordinate (C) at (.5,0);
		\coordinate (D) at (-.5,-1);
		\coordinate (E) at (-1.5,-1);
		\coordinate (F) at (.5,-2);

		\draw (B)  node [above] {$\vb{A1}$};
		\draw (C)  node [right] {$\vb{A2}$};
		\draw (D)  node [above left] {$\vb{A3}$};
		\draw (E)  node [left]  {$\vb{A4}$};
		\draw (F)  node [right] {$\vb{A5}$};

		\draw[thick, red, -]    (A)-- (B);
		\draw[thick, red, -]    (B)-- (C);
		\draw[thick, red, -]    (C)-- (D);
		\draw[thick, red, -]    (D)-- (E);
		\draw[thick, red, -]    (E)-- (F);
		
		\draw[thick, red, ->]   (A) -- ($.8*(A)+.2*(B)$);
		\draw[thick, red, ->]   (B) -- ($.7*(B)+.3*(C)$);
		\draw[thick, red, ->]   (C) -- ($.8*(C)+.2*(D)$);
		\draw[thick, red, ->]   (D) -- ($.7*(D)+.3*(E)$);
		\draw[thick, red, ->]   (E) -- ($.8*(E)+.2*(F)$);	
			
		\coordinate (A') at (4,-3);
		\coordinate (B') at (4,1);
		\coordinate (C') at (5,-2);
		\coordinate (D') at (2.5,0);
		\coordinate (E') at (3.6,-.5);
		\coordinate (F') at (4.25,-.4);

		\draw (B')  node [above] {$\vb{B1}$};
		\draw (C')  node [right] {$\vb{B2}$};
		\draw (D')  node [above] {$\vb{B3}$};
		\draw (E')  node [above]  {$\vb{B4}$};
		\draw (F')  node [below] {$\vb{B5}$};

		\draw[thick, blue, -]    (A')-- (B');
		\draw[thick, blue, -]    (B')-- (C');
		\draw[thick, blue, -]    (C')-- (D');
		\draw[thick, blue, -]    (D')-- (E');
		\draw[thick, blue, -]    (E')-- (F');
		
		\draw[thick, blue, ->]   (A') -- ($.8*(A')+.2*(B')$);
		\draw[thick, blue, ->]   (B') -- ($.8*(B')+.2*(C')$);
		\draw[thick, blue, ->]   (C') -- ($.8*(C')+.2*(D')$);
		\draw[thick, blue, ->]   (D') -- ($.7*(D')+.3*(E')$);
		\draw[thick, blue, ->]   (E') -- ($.6*(E')+.4*(F')$);
	
		\foreach \n in {B,C,D,E,F,B',C',D',E',F'}
			\node at (\n)[circle,fill, inner sep=1pt]{};	
		
\end{tikzpicture}
}	 
\caption{\centering	Two systems with 4 mirrors that are, by Proposition~\ref{propo:winding_num}. not in the same path connected component of $\mathcal{A}_4$.}
\end{figure}

\vspace{2mm}

\begin{proofnosquare}[Proof of Proposition~\ref{propo:winding_num}]
	See Appendix~\ref{sec:appendix_proofs} and \ref{sec:appendix_theorem_winding}.
\end{proofnosquare}

For telescopes with $n$ mirrors and an object at infinity, we can, without loss of generality, choose $a_0=\left(\begin{matrix}0\\-1\end{matrix}\right)$ and $a_1=\left(\begin{matrix}0\\0\end{matrix}\right)$ (via a translation, an homothety and a rotation). As the object is at infinity  we consider that the whole infinite half straight line $(a_0;a_1]$ is in the light and we do not consider $a_0$ in the constraints. 

\vspace{5mm}

\begin{definition}
\label{def:poly_telesc}

For telescopes, the sets we will study in the rest of this paper are the following:

\begin{align*}
	\widetilde{\mathcal{A}_n} &= \biggl\{\bfa \in (\mathbb{R}^2)^{n},\ 
	\Wedge\limits_{\substack{2\leqslant j\leqslant n+1}} \!\left(a_j\not\in (a_0;a_{1}]\right)\ \textrm{ and }
	\Wedge\limits_{\substack{1\leqslant i\leqslant n \\ 1\leqslant j\leqslant n+1\\j \notin \{i-1,i,i+1,i+2\}}} \!\left(a_j\not\in [a_i;a_{i+1}]\right)\\
	&\quad \quad \quad  \text{ and }  \Wedge\limits_{0\leqslant k\leqslant n-1 } a_{k+2}\not \in (a_k;a_{k+1})\biggr\}\\	
\end{align*}
\begin{align*}
	\widetilde{\mathcal{B}_n} &= \biggl\{\bfa \in (\mathbb{R}^2)^{n},\ 
	\Wedge\limits_{\substack{0\leqslant i\leqslant n \\ 1\leqslant j\leqslant n+1\\j \notin \{i,i+1\}}} \!\left(a_j\not\in (a_i;a_{i+1})\right)\biggr\}\\	
\end{align*}

The proofs of Proposition~\ref{propo:AB_open_alge}. and Proposition~\ref{propo:AB_ineq}. can be adapted to obtain $\lvert C\!\!\!C(\widetilde{\mathcal{A}_n}) \rvert \leqslant \lvert C\!\!\!C(\widetilde{\mathcal{B}_n}) \rvert$.
Propositions \ref{propo:sense_rota}. to \ref{propo:winding_num}. still hold considering $[a_0;a_1]$ as $(a_0;a_1]$.
\label{def:AnBnSemialgebraic_telecopes}
\end{definition}

In hindsight, it will be seen that the four invariants exhibited through Propositions \ref{propo:sense_rota}. to \ref{propo:winding_num}. are sufficient to characterize $C\!\!\!C(\widetilde{\mathcal{A}_4})$ with four mirrors, and that only the first three are sufficient for $C\!\!\!C(\widetilde{\mathcal{A}_3})$.
Hence, the semialgebraic expression of $\widetilde{\mathcal{A}_n}$ and $\widetilde{\mathcal{B}_n}$ are given by the following proposition.
\begin{proposition}
	\label{prop:AnBnSemialgebraic_telecopes}

Let be $a_i=\left(x_i,\  y_i\right)$ and
\begin{equation*}
	\begin{aligned}
	\hat{P}(a_1,a_2,a_3) &= x_1y_2-x_1y_3+x_2y_3-x_2y_1+x_3y_1-x_3y_2  \in\mathbb{Q}[x_1,x_2,x_3,y_1,y_2,y_3],\\
	\hat{R}(a_1,a_2,a_3) &= (x_2-x_1)(x_3-x_1)+(y_2-y_1)(y_3-y_1)\in\mathbb{Q}[x_1,x_2,x_3,y_1,y_2,y_3].
	\end{aligned}
\end{equation*}
For $i,j,k \in [\![0,n]\!]$ we adapt the definition of $P_{(i,j,k)}$ (resp. $R_{(i,j,k)}$) made in the proof of Proposition~\ref{propo:AB_open_alge}  for $\widetilde{\mathcal{B}_n}$ (resp. $\widetilde{\mathcal{A}_n}$).
We set $x_1=y_1=0$, then $P_{(i,j,k)},R_{(i,j,k)}\in\mathbb{Q}[x_2,...,x_n,y_2,...,y_n]$ are given by: 
\begin{equation*}
\begin{split}
P_{(0,1,k)} &= x_k \quad with \quad k  \in [\![2,n+1]\!]\; ,\\
P_{(i,j,k)}\; &= \hat{P}(a_i,a_j,a_k) \quad with \quad  \begin{cases}i,j,k  \in [\![1,n+1]\!] \quad and \quad i<j<k \\ \min(j-i,k-j)=1 \end{cases},\\
R_{(0,1,k)} &= -Y_k \quad with \quad k  \in [\![2,n+1]\!]\; ,\\
R_{(i,j,k)}\; &= \hat{R}(a_i,a_j,a_k) \quad with \quad\begin{cases}  i,j,k  \in [\![1,n+1]\!] \quad\textrm{distincts} \\ \lvert i-j\rvert = 1\end{cases}.
\end{split}
\end{equation*}

The conditions impose that a pair of points define a flux, thus are consecutive. As a permutation of $(i,j,k)$ only changes the sign of $\hat{P}(X_i,X_j,X_k)$, hence we can impose the order.
The set of all the distincts $P_{(i,j,k)}$  for a telescope with $n$ mirrors will be written $\mathcal{P}_n$, and we define 
\[
	Q_{n} = \prod\limits_{P\in \mathcal{P}_n} P\in \mathbb{Q}[x_2,...,x_{n+1},y_2,...,y_{n+1}].\]
Thus $\widetilde{\mathcal{B}_n}$ expresses as a Zariski open subset of  $\mathbb{R}^{2n}$:
\[\widetilde{\mathcal{B}_n} = \big\{ (x_2,...,x_{n+1},y_2,...,y_{n+1})\in\mathbb{R}^{2n},\ Q_n(x_2,...,x_{n+1},y_2,...,y_{n+1})\!\neq\! 0\big\}.
\]
Let us note that $deg\,Q_n=2C_n+n=(2n-3)n+2$ with $C_n=n(n-2)+1=\sharp\ \{P_{(i,j,k)}\in \mathcal{P}_n,\ (i,j)\neq(0,1)\}$.\\
And $\widetilde{\mathcal{A}_n}$ is a semialgebraic containing $\widetilde{\mathcal{B}_n}$:
\begin{align*}
	\widetilde{\mathcal{A}_n} &= \big\{ (a_2,...,a_{n+1})\in\mathbb{R}^{2n},\ \Wedge\limits_{\substack{ 2\leqslant j\leqslant n+1}} \left(\hat{P}(a_0,a_{1},a_j) \neq 0\right)\vee \left(\hat{R}(a_1,a_{0},a_j) < 0 \right)\\
	&\textrm{ and } \Wedge\limits_{\substack{1\leqslant i\leqslant n \\ 1\leqslant j\leqslant n+1\\j \notin \{i-1,i,i+1,i+2\}}} \!\left(\hat{P}(a_i,a_{i+1},a_j) \neq 0\right)\vee \left(\hat{R}(a_i,a_{i+1},a_j) < 0 \right)\vee \left(\hat{R}(a_{i+1},a_{i},a_j) < 0 \right)\\
	&\quad \quad \quad  \text{ and }  \Wedge\limits_{0\leqslant k\leqslant n-1 } \hat{P}(a_k,a_{k+1},a_{k+2})\neq 0\big\}.
\end{align*}
\end{proposition}
\begin{proof}
	Straightforward from Definition~\ref{def:AnBnSemialgebraic_telecopes} and proof of Proposition~\ref{propo:AB_open_alge}.
\end{proof}
\vspace{2mm}

\subsection{ \label{subsec:goffa}\textit{GOFFA} nomenclature}

Based on the studied invariants, we can create a Geometrical OFF Axis (GOFFA) nomenclature and assign a name to each class of systems.

The name of a system is constructed recursively on its fluxes. No letter is assigned to the first flux, then if a flux does not cross any previous flux we assign it the letter \textit{V}, else we assign it the letter \textit{X}. The letter are chosen as there shapes mimics that of the rays. In both cases we note in subcript of the letter the sense of rotation relative to the previous flux, ie. the sign of $(a_{i+2}-a_i) \wedge (a_{i+2}-a_{i+1})$ (\textit{A} for anti-clockwise and \textit{C} for clockwise). Thus if we remove the symmetry regarding the first fluxes, all names begin by $V_A$.

\vspace{2mm}

For systems with four mirrors, if the flux is crossing we also precise the indices of the fluxes that were crossed, as well as the sense of crossing, ie. the sign of $(a_{i+1}-a_i)\wedge(a_{j+1}-a_j)$ in case of intersection between $[a_{i+1}-a_i]$ and $[a_{j+1}-a_j]$.

\vspace{2mm}

Finaly, when needed, we write if the k-th surface is in a bounded or unbounded connected componant of the plane, \textit{b} for bounded and \textit{u} for unbounded in superscript. In practice there are only two pairs of systems with 4 mirrors for which this specification is needed : $V_AX_{A,0^+}\!V_CX^b_{A,0^-}$ , $V_AX_{A,0^+}\!V_CX^u_{A,0^-}$ and $V_AX_{A,0^+}\!V_AX^b_{C,0^-}$ , $V_AX_{A,0^+}\!V_AX^u_{C,0^-}$. In the rest of this article we will specify the latter only for those 4 families.

\vspace{5mm}

\begin{theorem}
From the foregoing:

\begin{center}Two systems of $\mathcal{A}_3$ (resp. $\mathcal{A}_4$) belong to the same connected component \\if and only if they share the same GOFFA name.
\end{center}
\end{theorem}

\subsection{Telescopes classification}\vspace{2mm}

\paragraph{Application.}

With Proposition~\ref{prop:AnBnSemialgebraic_telecopes}, we have 
\[
	Q_2=x_2x_3P_{(1,2,3)}=x_2x_3(x_2y_3-x_3y_2).
	\]
	 Hence $deg\,Q_2 = 4$ and \algo on $\widetilde{B}_2$ gives 8 points and we find by linear path (cf. section~\ref{sec:materiel_method}) the 6 connected components (3 with axial symmetry regarding the first flux) coresponding to $V_AV_A$ , $V_AV_C$ and $V_CX_{A,0}^2$.
 Thus : 
\begin{equation*}
\boxed{\lvert C\!\!\!C(\widetilde{\mathcal{A}}_2) \rvert = 6} \quad.
\end{equation*}
By adding a mirror we get $$Q_3=x_2x_3x_4P_{(1,2,3)}P_{(1,2,4)}P_{(2,3,4)}.$$
 We obtain $deg\,Q_3 = 11$ and \algo on the set $\widetilde{\mathcal{B}_3}$ gives  192 points (96 with symmetry). We find by the linear path heuristic (cf. section~\ref{sec:materiel_method}) on $\widetilde{\mathcal{A}_3}$, 32 connected components (16 with symmetry), hence :
\begin{equation*}
\boxed{\lvert C\!\!\!C(\widetilde{\mathcal{A}}_3) \rvert = 32} \quad.
\end{equation*}
Similarly, $$Q_4=x_2x_3x_4x_5P_{(1,2,3)}P_{(1,2,4)}P_{(1,2,5)}P_{(2,3,4)}P_{(2,3,5)}P_{(2,4,5)}P_{(3,4,5)}.$$
We obtain $deg\,Q_4 = 22$ and \algo on $\widetilde{B}_4$ returns  25 920 points (12 960 with symmetry). We join them via the linear path heuristic to obtain an upper bound of 320 components (160 with symmetry). With well chosen intermediate points (cf. section~\ref{sec:materiel_method} for materials and details) we obtain 144 different components (with symmetry) which all have a distinct set of invariants, hence: 
\begin{equation*}
\boxed{\lvert C\!\!\!C(\widetilde{\mathcal{A}}_4) \rvert = 288} \quad.
\end{equation*}

\vspace{-0cm}
\begin{figure}[H]\centering
\begin{subfigure}{\textwidth}
\includegraphics[width=\linewidth]{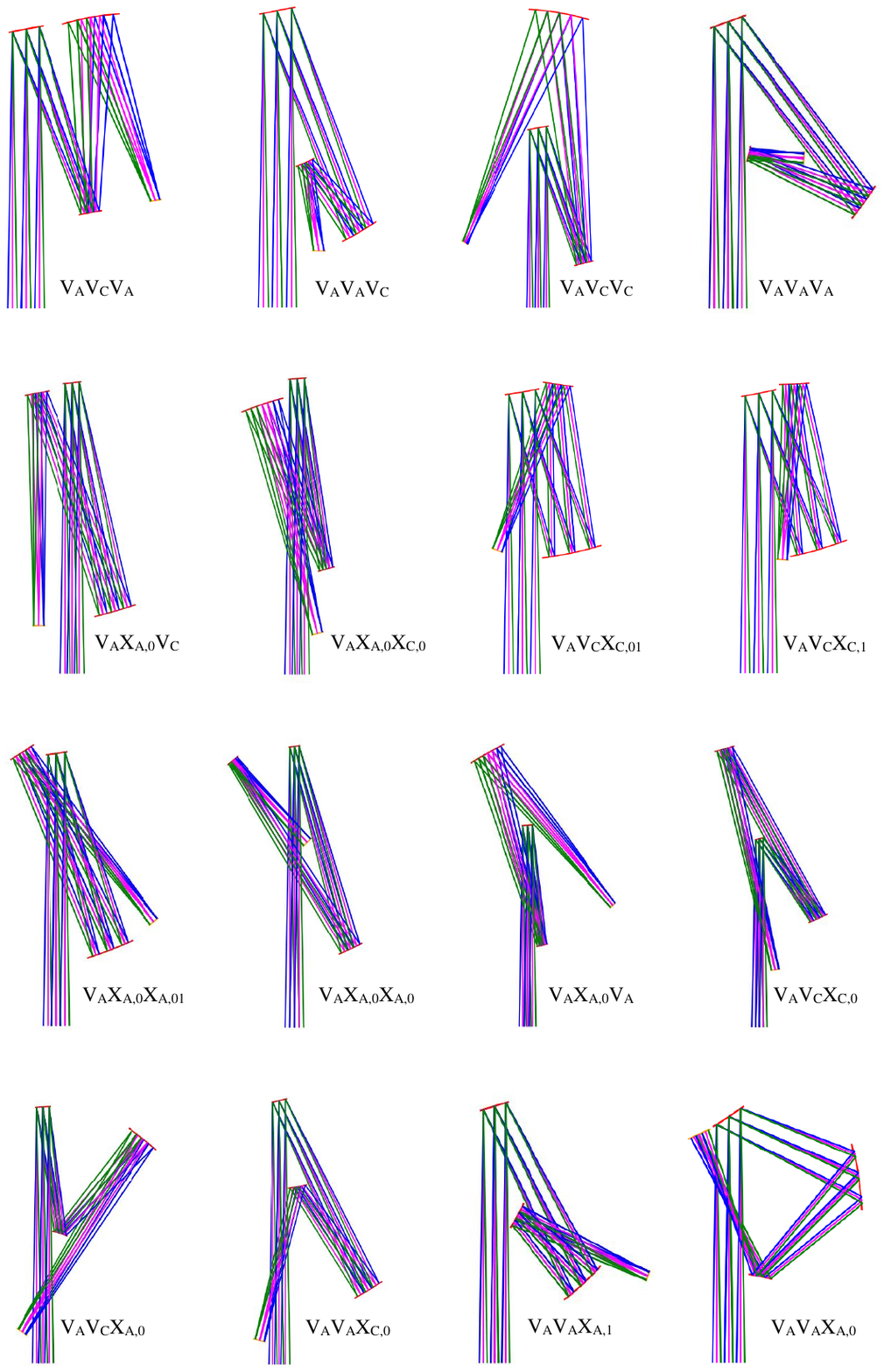}
\end{subfigure}
\caption{Off axis classification: all topologies of 3 mirror telescopes}
\label{flux3M}
\end{figure}

\pagebreak

\phantom{ }

The four figures below depict the entire set of 144 classes of unfolded 4-mirror systems whose names start with $V_A$. Only one ray is represented. The remaining classes consist of systems that are symmetric with respect to the first flux, and their names are obtained by simply replacing their first two characters with $V_C$.

\begin{figure}[H]\centering
\begin{subfigure}{10cm}
\includegraphics[width=10cm]{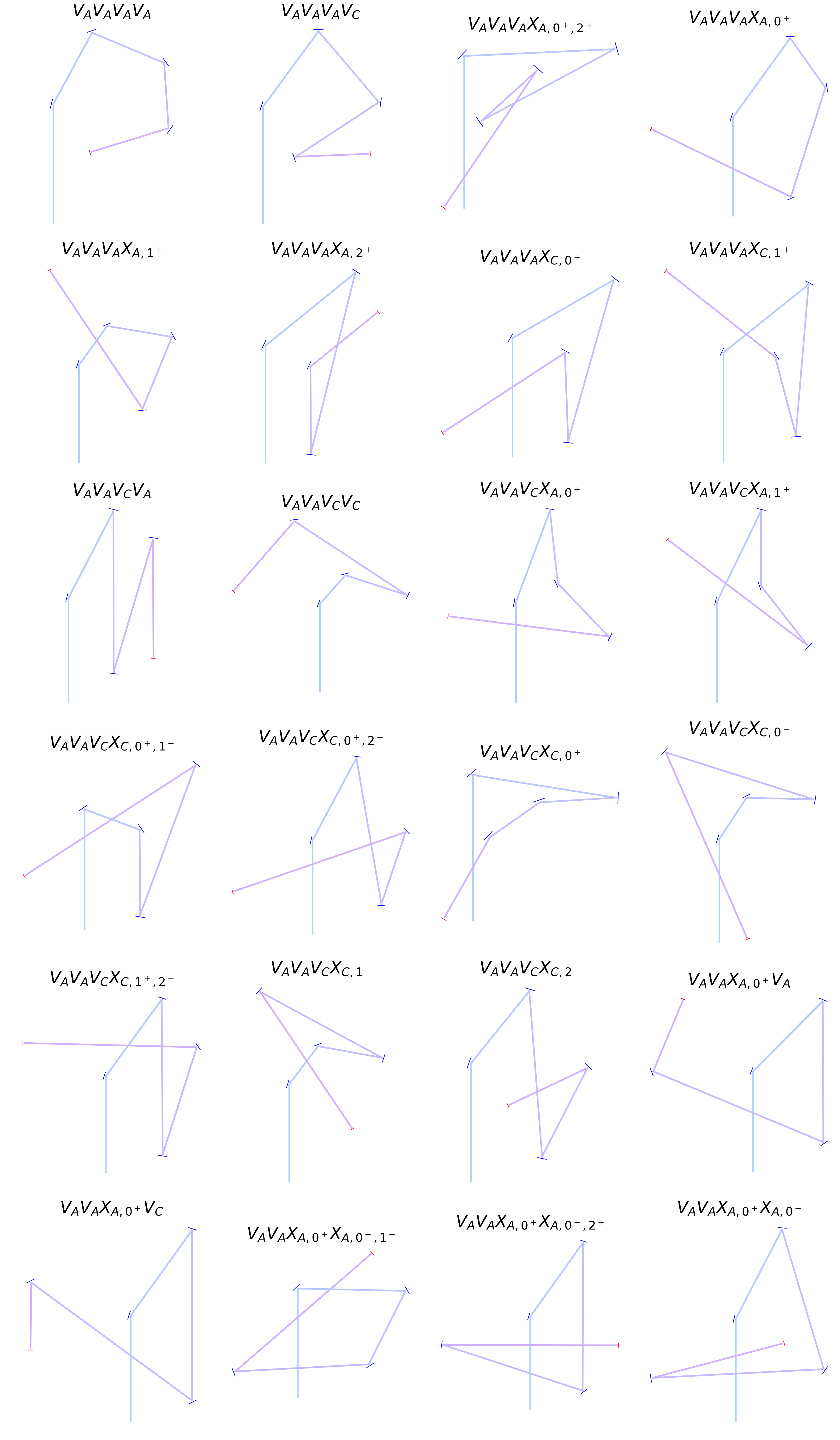}
\end{subfigure}
\caption{Off axis classification: all topologies of 4 mirrors telescopes (1 on 6)}
\label{flux4M}
\end{figure}

\begin{figure}[H]\centering
\begin{subfigure}{11cm}
\includegraphics[width=11cm]{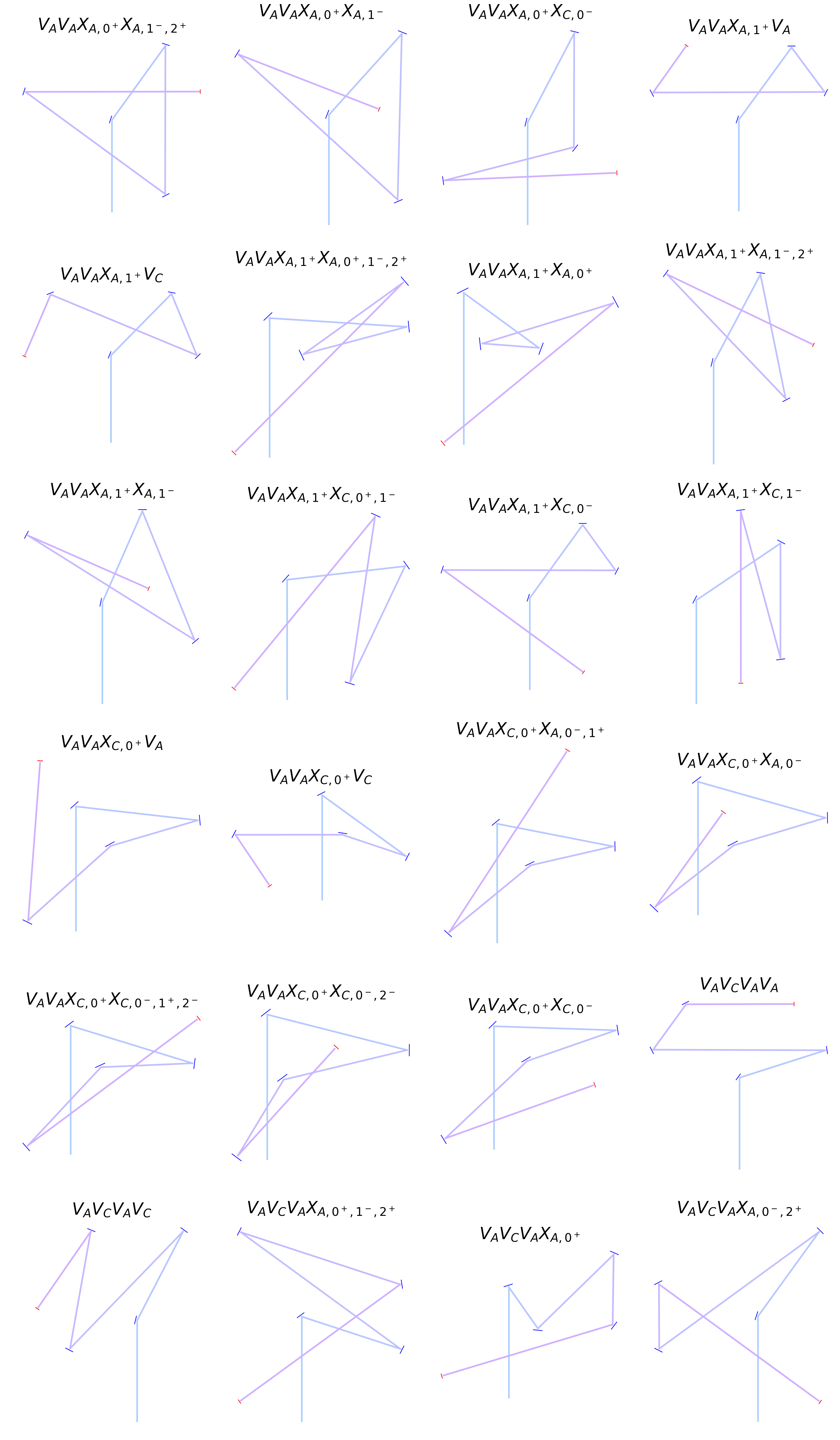}
\end{subfigure}
\caption{Off axis classification: all topologies of 4 mirrors telescopes (2 on 6)}
\label{flux4M}
\end{figure}

\begin{figure}[H]\centering
\begin{subfigure}{11cm}
\includegraphics[width=11cm]{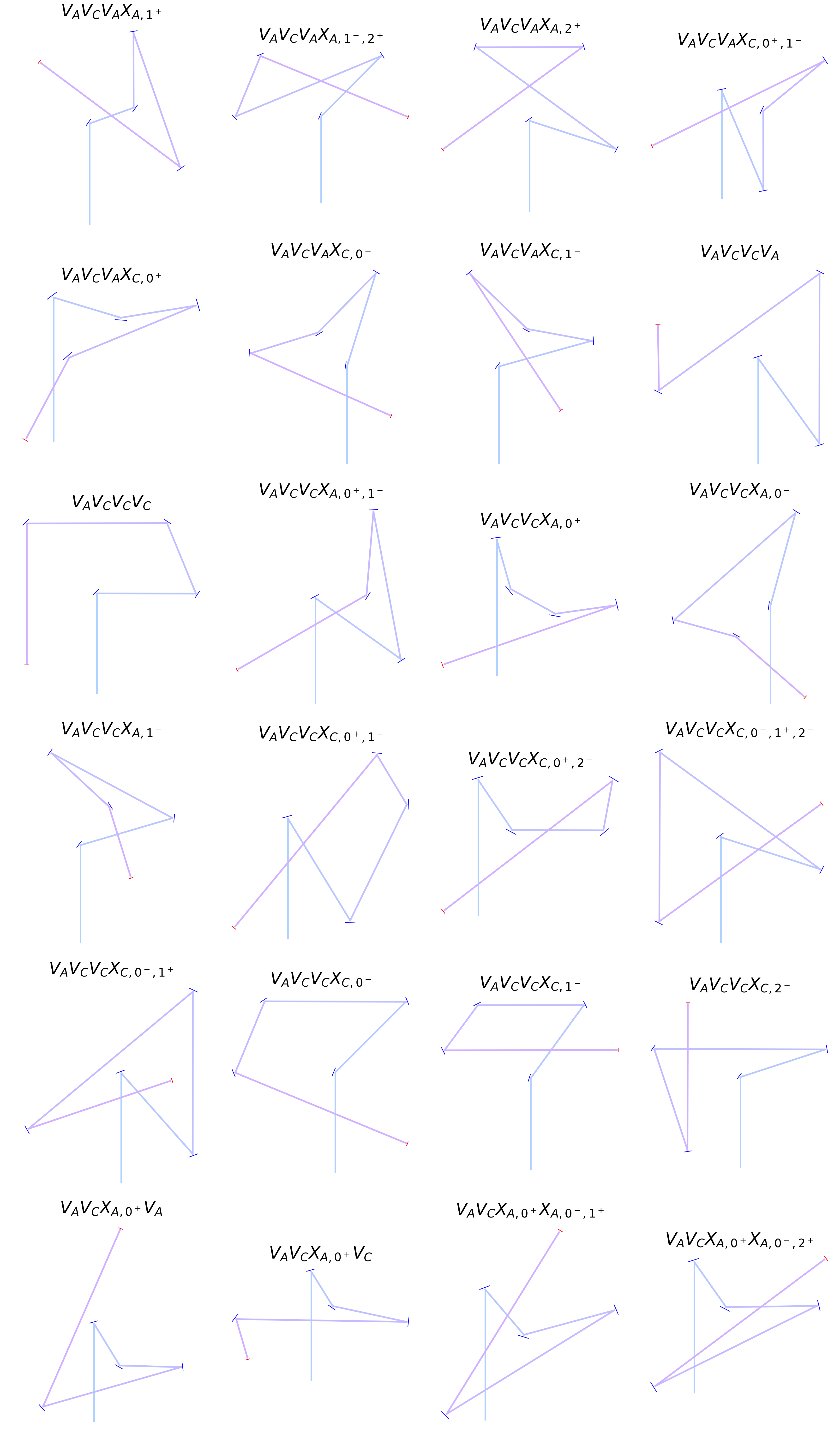}
\end{subfigure}
\caption{Off axis classification: all topologies of 4 mirrors telescopes (3 on 6)}
\label{flux4M}
\end{figure}

\begin{figure}[H]\centering
\begin{subfigure}{11cm}
\includegraphics[width=11cm]{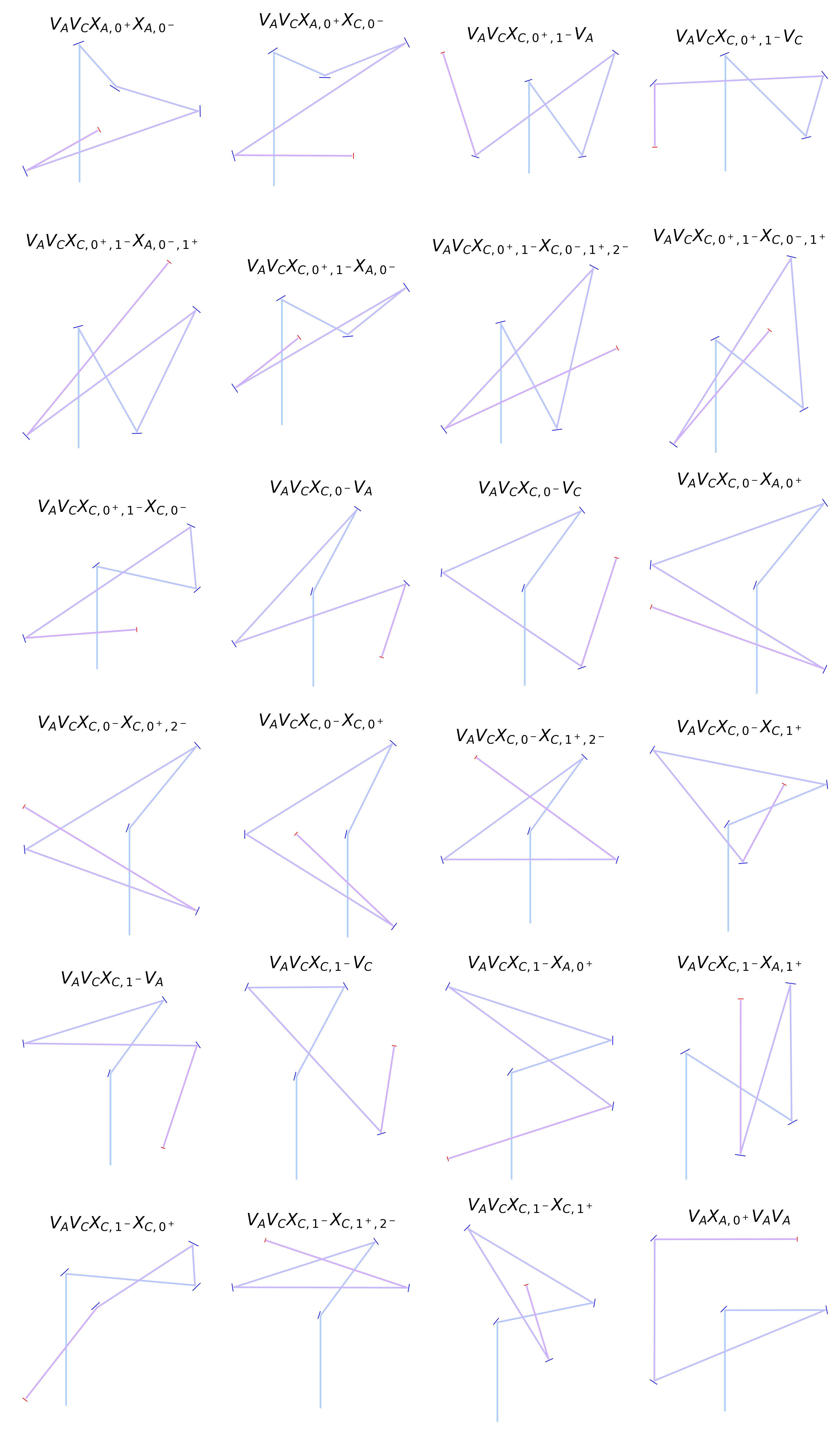}
\end{subfigure}
\caption{Off axis classification: all topologies of 4 mirrors telescopes (4 on 6)}
\label{flux4M}
\end{figure}

\begin{figure}[H]\centering
\begin{subfigure}{11cm}
\includegraphics[width=11cm]{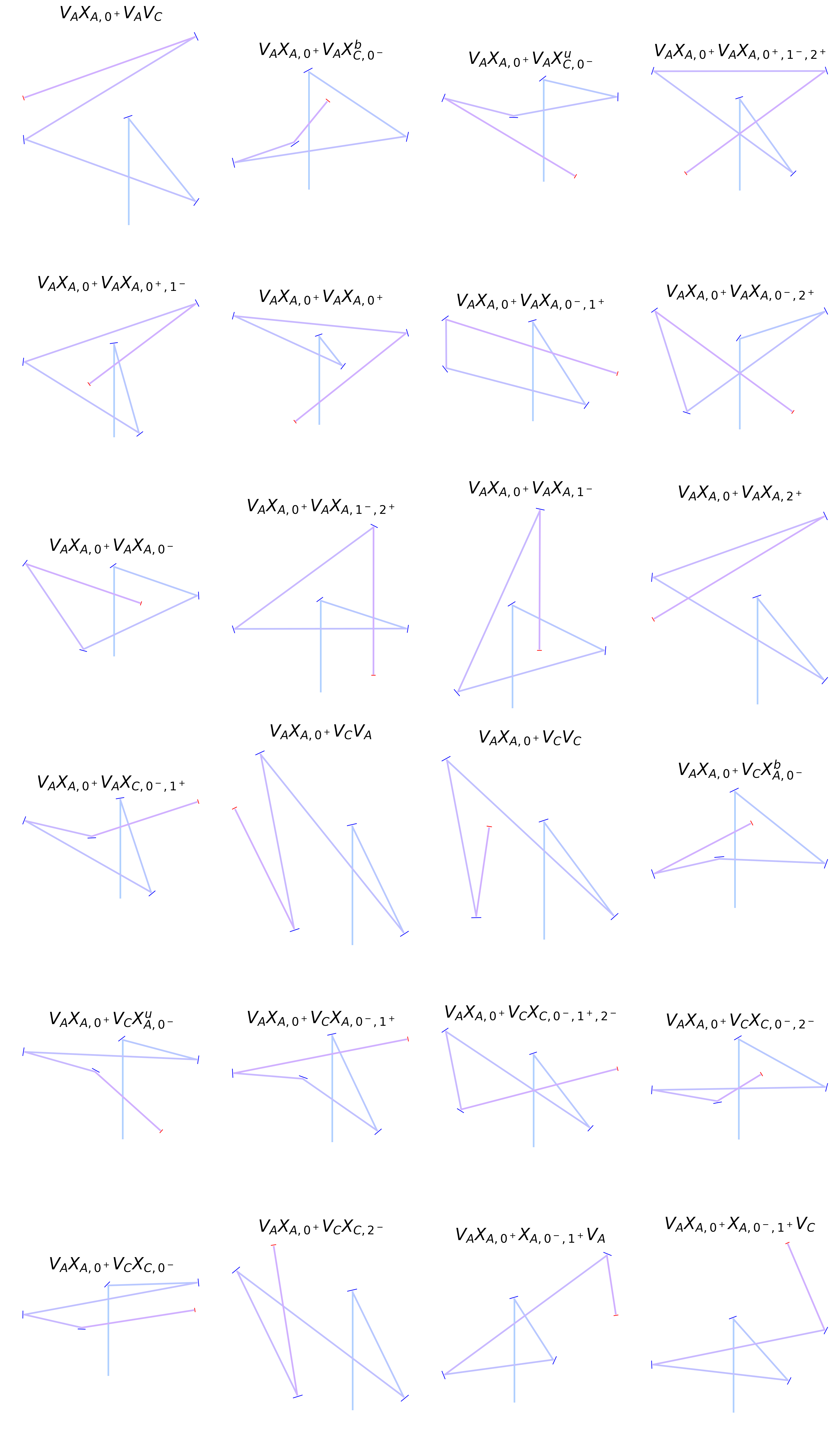}
\end{subfigure}
\caption{Off axis classification: all topologies of 4 mirrors telescopes (5 on 6)}
\label{flux4M}
\end{figure}

\begin{figure}[H]\centering
\begin{subfigure}{11cm}
\includegraphics[width=11cm]{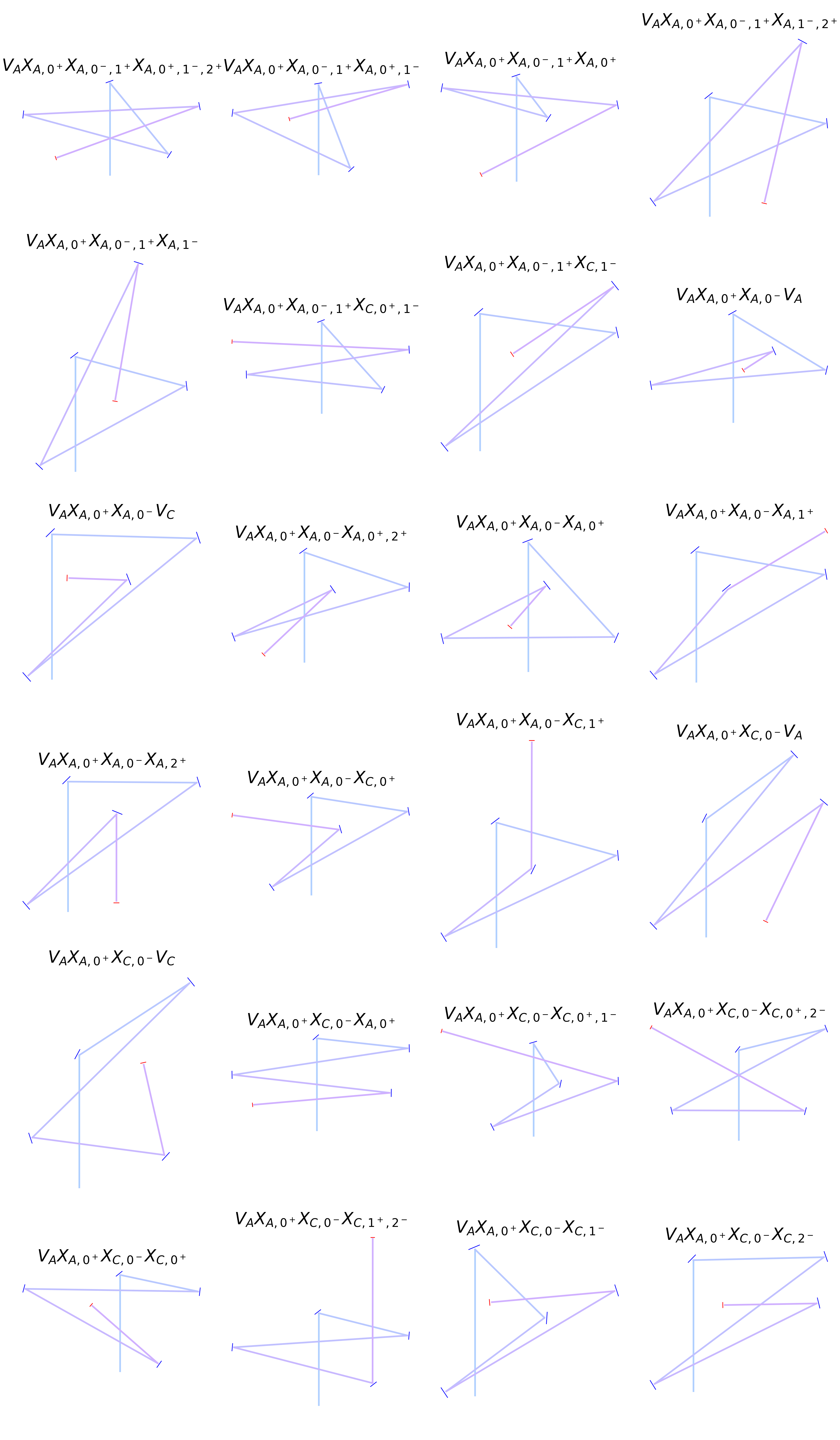}
\end{subfigure}
\caption{Off axis classification: all topologies of 4 mirrors telescopes (6 on 6)}
\label{flux4M}
\end{figure}

The proof of the previous results being constructive, we not only obtain the number of families but also a representative for each connected component. Thus, given a cost function corresponding to certain optical needs, and by imposing a high cost on obscuration, we achieve, through gradient descent (or another optimization algorithm), a local minimum on each connected component. Selecting the best system among the obtained solutions allows for comprehensiveness while obtaining a unobscured system.

All classes of solutions can be implemented both in practice and in simulation by reducing the diameters of the entrance pupils to zero. Of course, some families inherently exhibit better optical performance than others, and existing as well as future publications will showcase the best systems for a given set of specifications \cite{Zhu:21}. 

The tilt of the mirrors deviates the system from the first order approximation, and a strong tilt often results in poor optical performance . Thus, it is possible to exclude families that cannot approach a sum of tilts equal to 0. With three mirrors, the only such family of this is $V_AV_AX_{A,0}$ (and its symmetrical).

Time comlplexity is the reason why the resolution could not be made for systems with 5 mirrors (indeed the time complexity of  \algo is doubly exponential in the dimension of the unknown \cite{pollackBasuRoy}). 

If the system is no longer constrained to be symmetrical with respect to a plane, it is then possible to extend the system in three dimensions to eliminate obscuration. 
In this case, a dimension argument can be used to demonstrate that there exists a unique connected component (see Appendix~\ref{sec:appendix_3D}). However, in practice, the total volume of the system is constrained, and taking into account the thickness of the ray also leads to the emergence of multiple solution of unobscured families.

%% file: inputs/methods.tex
\subsection{Cylindrical Algebraic Decomposition}

The invariants can give us a lower bound of $\lvert C\!\!\!C(\widetilde{\mathcal{A}}_n) \rvert$ given that we have representatives from different components. 
Given  \algo (Cylindrical Algebraic Decomposition \cite{Collins:75,pollackBasuRoy}) computing at least one point by connected component, applied on the following polynoms describing $\widetilde{\mathcal{B}}_n$, will give us a first upper bound as well as representatives from every components.
 Finaly we will exhibit continuous paths (cf. section~\ref{sec:materiel_method}) to get the upper bound given by  \algo down to the lower bound given by the invariants.

 For example, the four mirrors  instances output of \algo were obtained in Mathematica \cite{Wolfram:22} computing the following command calling the function \textit{SemialgebraicComponentInstances}  (we only compute \verb|>0| as the other instances are obtained by axial symmetry with respect to the first flux): \\
 \verb|SemialgebraicComponentInstances[x2 * x3 * x4 * x5 | \\
 \phantom{\qquad}\verb|* (x2*y3 - x3*y2) * (x2*y4 - x4*y2) * (x2*y5 - x5*y2)| \\
 \phantom{\qquad}\verb|* (x3*y4 - x4*y3) * (x4*y5 - x5*y4)| \\
 \phantom{\qquad}\verb|* (x2*y3 - x2*y4 - x3*y2 + x3*y4 + x4*y2 - x4*y3)| \\
 \phantom{\qquad}\verb|* (x2*y3 - x2*y5 - x3*y2 + x3*y5 + x5*y2 - x5*y3)| \\
 \phantom{\qquad}\verb|* (x2*y4 - x2*y5 - x4*y2 + x4*y5 + x5*y2 - x5*y4)| \\
 \phantom{\qquad}\verb|* (x3*y4 - x3*y5 - x4*y3 + x4*y5 + x5*y3 - x5*y4) >0 | \\
 \phantom{\qquad}\verb|,{x2,x3,x4,x5, y2, y3,y4 ,y5}]| \\
 
\label{sec:materiel_method}
\subsection{Linear Path Heuristic on instances found solutions}

Even if the connected components are not convex, we try straight linear paths for every pair of solutions (obtained by \algo) that could be in the same connected components (based on the GOFFA nomenclature).

\vspace{2mm}

With $U,V\in \widetilde{\mathcal{A}}_n$ given by \algo, we introduce the path $T(\lambda) = (1-\lambda)U+\lambda V $; we test for $P\in \mathcal{P}_n$, if $P\!\circ\! T \in \mathbb{R}[\lambda]$ does not vanishes over $[0,1]$ . We can hope the upper bound obtained reaches the lower bound given by the 
GOFFA nomenclature.

\vspace{2mm}

In the case where $\exists P_{(i,j,k)}\in \mathcal{P}_n , \exists \lambda_0 \in [0,1], P\!\circ\! T(\lambda_0)=0$ :
\begin{itemize}
\setlength{\itemindent}{+5mm}
\item If $(i,j,k)$ are consecutive, $P\!\circ\! T(\lambda_0) \notin \widetilde{\mathcal{A}}_n$ .
\item Else, we check the signs of $R_{(i,j,k)}(\lambda_0)$ and $R_{(j,i,k)}(\lambda_0)$, if they are both positive $P\!\circ\! T(\lambda_0) \notin \widetilde{\mathcal{A}}_n$.
\end{itemize}

Checking this for all $P$ in $\mathcal{P}_n$ is sufficient to show that $P\!\circ\! T([0,1]) \subset \widetilde{\mathcal{A}}_n$, that is $U,V$ are in the same path connected component of $\widetilde{\mathcal{A}}_n$.

\subsection{Intermediate points for four mirrors}

The only two pairs of systems that need the winding number criteria to be distinguished are $V_AX^2_{A,0^+}V_CX^4_{A,0^-},V_AX^2_{A,0^+}V_AX^4_{C,0^-}$\\
The folowing set is sufficient to group the remaining families that were not linked by straight path :

\noindent$V_AV_CV_AX^4_{A,0^+,1^-,2^+}:[20,20,41,-2,0,1,1,\frac{-1}{16}]$\\ 
$V_AV_CV_CX^4_{C,0^-,1^+}:[2,\frac{5}{2},-1,\frac{1}{2},0,1,\frac{-1}{4},\frac{1}{10}]$\\ 
$V_AV_CV_AX^4_{C,1^-}:[3,\frac{31}{32},2,\frac{3}{4},0,1,6,\frac{-1}{32}]$\\
$V_AV_CX^3_{C,1^-}X^4_{C,1^+,2^-}:[1,\frac{-3}{2},\frac{7}{2},-2,0,\frac{31}{160},\frac{-11}{32},\frac{5}{16}]$\\ 
$V_AX^2_{A,0^+}X^3_{A,0^-,1^+}X^4_{A,0^+,1^-,2^+}:[1,-9,\frac{23}{10},-2,0,-1,\frac{5}{32},\frac{-5}{16}]$\\ 
$V_AX^2_{A,0^+}X^3_{C,0^-}X^4_{C,1^+,2^-}:[1,-9,\frac{36}{25},-2,0,-1,\frac{-9}{400},\frac{1}{16}]$\\ 
$V_AX^2_{A,0^+}V_AX^4_{A,1^-,2^+}:[1,-9,\frac{-33}{20},\frac{29}{20},0,-1,\frac{9}{128},\frac{-5}{128}]$\\ 
$V_AV_CX^3_{C,0^-}X^4_{C,1^+}:[1,-9,41,\frac{1}{2},0,1,\frac{-2313839}{376509},\frac{1}{100}]$\\ 
$V_AV_CV_CX^4_{C,0^-,1^+,2^-}:[1,-2,\frac{-67}{50},\frac{44}{25},0,1,\frac{-19}{320},\frac{1}{20}]$\\ 
$V_AV_CX^3_{C,0^-}X^4_{C,0^+,2^-}:[1,-9,\frac{11}{5},-2,0,1,\frac{-17619}{51041},\frac{1229}{4096}]$\\ 
$V_AX^2_{A,0^+}V_CX^4_{A,0^-,1^+}:[5,\frac{-1}{10},-5,3,0,\frac{-1}{5},\frac{-39}{32},\frac{5}{8}]$\\ 
$V_AV_AX^3_{A,0^+}X^4_{A,1^-}:[1,4,-1,\frac{1}{2},0,-1,\frac{15}{64},\frac{-1}{10}]$\\ 
$V_AV_CV_AX^4_{C,0^+,1^-}:[5,1,6,-2,0,\frac{1}{10},\frac{39}{32},\frac{-1}{2}]$\\
$V_AV_AX^3_{C,0^+}X^4_{A,0^-}:[3,\frac{1}{4},-1,\frac{1}{10},0,-1,-3,\frac{-1}{2}],[3,\frac{1}{4},\frac{-1}{10},\frac{1}{10},0,-1,-3,\frac{-1}{2}]$\\ 
$V_AX^2_{A,0^+}X^3_{A,0^-}X^4_{A,0^+}:[1,-3,\frac{1}{10},\frac{-1}{2},0,-1,\frac{-1}{32},\frac{-23}{64}],[1,-1,\frac{1}{10},\frac{-1}{2},0,-1,\frac{-1}{8},\frac{-23}{32}]$\\
\phantom{\qquad\qquad\qquad\qquad}$[1,\frac{-47}{20},\frac{1}{10},-2,0,\frac{-85}{256},\frac{-1}{32},\frac{-19}{64}],[1,\frac{-47}{20},\frac{1}{10},\frac{-101}{1000},0,\frac{-85}{256},\frac{-1}{32},\frac{-37}{640}]$\\

\pagebreak

\phantom{ }